\documentclass[preprint]{elsarticle}
\biboptions{sort&compress}

\usepackage{graphicx,psfrag}
\usepackage{amsmath,amssymb}
\usepackage[list=true, font=normal, 
labelformat=brace, position=top]{subcaption}
\usepackage{algorithm,algorithmic}
\usepackage{lineno,hyperref}
\modulolinenumbers[5]

\newcommand{\sty}[1]{\mbox{\boldmath $#1$}}

\newcommand{\fu}{\sty{ u}}
\newcommand{\ff}{\sty{ f}}

\newcommand{\fv}{\sty{ v}}
\newcommand{\fn}{\sty{ n}}

\newcommand{\ft}{\sty{ t}}
\newcommand{\fx}{\sty{ x}}
\newcommand{\fy}{\sty{ y}}
\newcommand{\fz}{\sty{ z}}
\newcommand{\fB}{\sty{ B}}
\newcommand{\fI}{\sty{ I}}
\newcommand{\fR}{\sty{ R}}
\newcommand{\fP}{\sty{ P}}
\newcommand{\fT}{\sty{ T}}
\newcommand{\lb}{\left(}
\newcommand{\rb}{\right)}

\newcommand{\sv}{\lb\begin{array}{cccccccccc}}
\newcommand{\ev}{\end{array}\rb}

\newcommand{\feps}{\mbox{\boldmath $\varepsilon $}}
\newcommand{\fsigma}{\mbox{\boldmath $\sigma$}}
\newcommand{\feta}{\sty{ \eta}}
\newcommand{\e}{^{(e)}}

\newcommand{\es}{^{*(e)}}
\newcommand{\eT}{^{(e)T}}

\newcommand{\Ass}{\overset{m}{\underset{e=1}{\boldsymbol{\mathrm A}}}}
\newcommand{\tr}[1]{{\rm tr  }( #1 )}
\newcommand{\norm}[1]{| #1 |}
\newcommand{\sign}[1]{{\rm sgn}\left( #1 \right)}
 
\begin{document}

\begin{frontmatter} 
\title{Model-free Data-Driven Computational Mechanics Enhanced by Tensor Voting}
\author[ifam]{Robert Eggersmann\corref{cor1}}
\ead{robert.eggersmann@ifam.rwth-aachen.de}
\author[icme]{Laurent Stainier}
\author[aero,hcm]{Michael Ortiz}
\author[ifam]{Stefanie Reese}
\cortext[cor1]{Corresponding author}
\address[ifam]{Institute of Applied Mechanics, RWTH Aachen University, Mies-van-der-Rohe-Str.1, D-52074 Aachen, Germany.}
\address[icme]{Institute of Civil and Mechanical Engineering, \'{E}cole Centrale de Nantes, 1 Rue de la No\"e, F-44321 Nantes, France.}
\address[aero]{Division of Engineering and Applied Science, California Institute of Technology, Pasadena, CA 91125, USA.}
\address[hcm]{Hausdorff Center for Mathematics, Rheinische Friedrich-Wilhelms-Universit\"at Bonn, Endenicher Allee 62, D-53115 Bonn, Germany.}

\begin{abstract}
The data-driven computing paradigm initially introduced by Kirchdoerfer \& Ortiz (2016)  \citep{kirchdoerfer2016data} is extended by incorporating locally linear tangent spaces into the data set.
These tangent spaces are constructed by means of the tensor voting method introduced by  Mordohai \& Medioni (2010) \citep{mordohai2010dimensionality} which improves the learning of the underlying structure of a data set.
Tensor voting is an instance-based machine learning technique which accumulates votes from the nearest neighbors to build up second-order tensors encoding tangents and normals to the underlying data structure.
The here proposed second-order data-driven paradigm is a plug-in method for distance-minimizing as well as entropy-maximizing data-driven schemes.
Like its predecessor  \citep{kirchdoerfer2016data}, the resulting method aims to minimize a suitably defined free energy over phase space subject to compatibility and equilibrium constraints. 
The method's implementation is straightforward and numerically efficient since the data structure analysis is performed in an offline step.
Selected numerical examples are presented that establish the higher-order convergence properties of the data-driven solvers enhanced by tensor voting for ideal and noisy data sets. 
\end{abstract}

\end{frontmatter}

\section{Introduction}\label{sec:intro}
\noindent
The accumulation, storage and analysis of data have become one of the most discussed scientific topics over the past decade.
Sometimes data is even mentioned to become the new currency.
In scientific computing, the topic of data usage and machine learning attracts more and more attention.
Based on the far-reaching developments in the experimental field and measurement technology, many methods are developed to make use of this data-richness.\\[0.3cm]
In solid mechanics we face data in the fundamental form of strains and stresses defining the phase space.
Traditionally, the aim is to find a phenomenological model that represents the data in the best way possible.
This very demanding task is tackled extensively and successfully. Nevertheless, many models reach a degree of complexity which is hard to handle and understand.
For this reason, also in solid mechanics, the use of machine learning techniques has started to become a central emphasis for research. 
Global machine learning techniques as e.g. artificial neural networks perform astonishingly well for many kinds of problems.
Unfortunately, they act like some kind of ``black box", especially for higher-dimensional cases, and it is difficult to assess the range of validity for these approximations.\\[0.3cm]
Another approach is concentrating on instance-based or memory-based learning where constitutive manifolds are constructed from nearest neighbors without a global approximation.
Ibanez et al. \citep{ibanez2018manifold} propose a manifold learning approach using locally linear embedding.
The data is mapped into a lower-dimensional space to identify key dependencies in the material behavior.
Regrettably, the method becomes less effective for nonlinear problems.
Therefore, in \citep{gonzalez2018kpca} a manifold is established based on kernel principal component analysis.
Here, the data is mapped into a space of higher dimensionality in order to linearize the data. \\[0.3cm]
In contrast, the data-driven computational mechanics paradigm \citep{kirchdoerfer2016data,conti2018data}, called from now on DDCM for brevity, incorporates the data in the computation directly.
The method is based on a nearest neighbors approach.
By reformulating the initial boundary value problem in terms of minimizing the distance between, on the one hand, the constraint set and, on the other hand, the material data set, a generalization towards the classical problem of solid mechanics is formulated.
The constraint set contains all members in phase space which fulfill balance and kinematic relations.
The material data set includes stress and strain data from experimental measurements.\\[0.3cm]
The new approach was complemented by the entropy-maximizing data-driven paradigm \citep{kirchdoerfer2017data}. 
The latter represents a robust extension of the earlier approach, the advantages of which are in particular visible in the context of noisy data sets with outliers.
Further extensions were developed for finite deformations \citep{nguyen2018data,platzer2019assessment}, dynamics \citep{kirchdoerfer2018data} and inelasticity \cite{eggersmann2019model}.
An inverse formulation of the DDCM to identify stresses from measured strain fields was initially formulated by Leygue et al. \citep{leygue2018data}.
Further examination of this approach was carried out in \citep{dalemat2019measuring,leygue2019non}.
A combination of the stress identification algorithm (\citep{leygue2018data}) and the data-driven solver (\citep{kirchdoerfer2016data}) was investigated by Stainier et al. \citep{stainier2019model}.
In \citep{korzeniowski2019comparison}, the DDCM method was applied and compared to a statistical finite element analysis.
Another solution scheme towards DDCM  was investigated in \citep{kanno2019mixed}, where the method was implemented using  mixed-integer programming.
\\[0.3cm]
One main problem of all data-driven approaches is the well-known curse of dimensionality.
In the case of solid mechanics, this results in the problem that data sampling for multidimensional cases is very challenging.
The higher the dimension, the sparser the distribution of data points becomes.
Thus, the central objective of the current work is to develop an extension to the minimum-distance (min-dist) and maximum-entropy (max-ent) data-driven computing paradigm, to be called data-driven tensor voting (ten-vote).\\[0.3cm]
The classical data-driven solution scheme developed so far uses nearest neighbors clustering only.
Nevertheless, considering a data set of a particular problem, it can be assumed that there is an underlying structure of the data.
That means data points are linked in some way and have a more or less clearly defined shape.
In the present work, the structure in data sets is used by means of pointwise tangent spaces, i.e., every data point is equipped with tangents and normals to a manifold which are computed in a tensor voting step according to Mordohai \& Medioni \citep{mordohai2010dimensionality}.
The learning of tangent spaces via tensor voting is a machine learning technique evaluating distances and directions to nearest neighbors.
It should be emphasized that it is also conceivable to choose other approaches to compute the here required tangents and normals to a manifold. 
See e.g. Zhang \& Zha \citep{zhang2004principal} where the local tangent space alignment (LTSA) algorithm is introduced, and Wang et al. \citep{wang2005adaptive} where the algorithm is investigated towards the optimal choice of the neighborhood size.
Nevertheless, the method chosen in the present paper has certain advantages which are pointed out in \mbox{Section \ref{sec:tensorvoting} }.\\[0.3cm] 
The data-driven computing paradigm is extended to a second-order scheme in the following way. 
First, the search of the closest material state in the material data set is performed as in the initial solution scheme.
Second, the closest state in the locally linear tangent spaces of corresponding local states is found. 
This procedure enables us to find additional states which are still very close to the original data.
Furthermore, it can be stated that the method is easy to implement and computationally inexpensive since the additional data analysis for the tensor voting is performed in an offline step.\\[0.3cm]
The paper is organized as follows.
In Section \ref{sec:datadriven}, theory and central equations of the data-driven computing paradigm (see \citep{kirchdoerfer2016data}) is summarized.
Section \ref{sec:tensorvoting} recapitulates the tensor voting method \citep{mordohai2010dimensionality} and shows how to deal with material data.
The proposed tensor voting procedure is then used in Section \ref{sec:ddtensorvoting} to enhance the data-driven computing paradigm.
Thereupon, two examples are investigated by numerical testing to show the applicability of the method.
In Section \ref{sec:trusses}, we investigate computations of truss structures for one-dimensional stress-strain data with noise-free and noisy data sets.
This is followed by further numerical tests of a two-dimensional continuum mechanical problem in Section \ref{sec:continuum}.
Concluding remarks and opportunities for further developments of the data-driven computing paradigm are presented in Section \ref{sec:conclusion}.
\section{Background: The data-driven computing paradigm}
\label{sec:datadriven}
\noindent
We start by summarizing the data-driven formulation of the discretized initial boundary value problem of elasticity, as proposed in \citep{kirchdoerfer2016data, conti2018data}. 
The system undergoes displacements $\fu =\{\fu_i\}_{i=1}^N$, with $\fu_i\in\mathbb{R}^{n_i}$ being the displacement vector at node $i$ with dimension $n_i$ at all nodes $i=1,...,N$, under the action of applied forces $\ff = \{\ff_i\}_{i=1}^{N}$. 
The vector $\ff_i\in\mathbb{R}^{n_i}$ denotes the nodal force vector. 
Stress and strain pairs $\{(\feps\e,\fsigma\e)\}_{e=1}^{m}$ define the local phase space with $\feps\e$, $\fsigma\e\in\mathbb{R}^{m_e}$ and $m_e$ being the dimension of stress and strain at material point $e=1,\dots,m$.
Regarding $\fz\e = (\feps\e,\fsigma\e)$ as a point in a local phase space $Z\e\in\mathbb{R}^{2m_e}$ we find $\fz=\{(\feps\e,\fsigma\e)\}_{e=1}^{m}$ as a point in the global phase space $Z=Z^{(1)}\times ... \times Z^{(m)}$.\\[0.3cm] 
The internal state of the system is constrained by the kinematic relation
\begin{equation}
\feps\e=\fB\e\fu\e,\hspace{1cm}\forall e=1,...,m\label{eq:constraint1}
\end{equation}
and the equilibrium condition
\begin{equation}
\Ass \{w\e\fB\eT\fsigma\e\}=\ff.\label{eq:constraint2}
\end{equation}
In the latter equations, the symbol $\Ass$ refers to the assembly of all elements and $\fB\e$ represents the standard strain-displacement operator for material point $e$. 
The prefactors $\{w\e\}_{e=1}^{m}$ denote the volumes of the elements.
The constraints \eqref{eq:constraint1} and \eqref{eq:constraint2} are material-independent and define a subspace
\begin{equation}
C=\{\fz\in Z:\eqref{eq:constraint1}\; \text{and}\; \eqref{eq:constraint2}\},
\end{equation}
which is denoted as constraint set. It consists of all compatible and equilibrated internal states.\\[0.3cm] 
Conventionally, the still missing connection between $\feps\e$ and $\fsigma\e$ (the material response) is given by a material law, i.e., functions of the general form
\begin{equation}	
	\fsigma\e=\hat\fsigma\e(\feps\e,...)\hspace{1cm}\forall e=1,...,m,\label{eq:materiallaw}
\end{equation} 
where $\hat\fsigma\e:\mathbb{R}^{m_e}\rightarrow\mathbb{R}^{m_e}$.\\[0.3cm] 
Here, a different point of view is followed.
The information about the material behavior under investigation is given by means of a material data set $D\e$, which classically consists of experimental measurements or data achieved from small scale simulations in the form of points $\fz\e=(\feps\e,\fsigma\e)\in Z\e$.\\[0.3cm] 
The data-driven reformulation of the initial boundary-value problem incorporates information about the material behavior directly in terms of the material data. 
In this way, the step of material modeling is bypassed completely. 
One class of data-driven problems consists of finding the internal state $\fz \in C$, which minimizes some distance $d$ to the global material data set $D=D^{(1)}\times ... \times D^{(m)}$. 
Therefore, a metric 
\begin{equation}
|\fz\e|_e=\Big(\frac{1}{2}\mathbb{C}\e\feps\e\cdot\feps\e+\frac{1}{2}\mathbb{C}^{(e)-1}\fsigma\e\cdot\fsigma\e\Big)^{1/2}\label{eq:distmetric}
\end{equation}
of the local phase space by means of some positive-definite and symmetric matrices $\{\mathbb{C}\e\}_{e=1}^{m}$ is introduced.
The corresponding distance
\begin{equation}
d_e(\fz\e,\fy\e)=|\fz\e-\fy\e|_e,\label{eq:norm}
\end{equation}
with $\fy\e,\fz\e\in Z\e$ induces a metric of the global phase $Z$ by means of the global norm
\begin{equation}
|\fz|=\Big(\sum_{e=1}^m w\e |\fz\e|_e^{2}\Big)^{1/2},
\end{equation}
with associated distance
\begin{equation}
d(\fz,\fy)=|\fz-\fy|,
\end{equation}
for $\fy,\fz\in Z$. 
The distance-minimizing data-driven problem is formulated by stating 
\begin{equation}
\underset{\boldsymbol y\in D}{\min}\;\underset{\boldsymbol z\in C}{\min}\; d(\fz,\fy) = \underset{\boldsymbol z\in C}{\min}\;\underset{\boldsymbol y\in D}{\min}\; d(\fz,\fy) \label{eq:minmin},
\end{equation}
i.e., the objective is to find the point $\fy \in D$ in the material data set that is closest to the constraint set $C$ of compatible and equilibrated internal states.
Equivalently, we aim at finding the compatible and equilibrated internal state $\fz \in C$ that is closest to the material data set $D$.\\[0.3cm] 
The benefit of the proposed data-driven solution scheme is that the local material data sets can be graphs, point sets or any arbitrary set in phase space. 
This property of the solution scheme will be exploited in the following work.
Additionally, it should be pointed out that the classical solution will be recovered if the local material data sets are chosen as
\begin{equation}
D\e=\{(\feps\e,\hat\fsigma(\feps\e))\},
\end{equation}
i.e. as graphs in $Z\e$ defined by the material law \eqref{eq:materiallaw}. 
Thus, the data-driven reformulation \eqref{eq:minmin} extends - and subsumes as special cases - the classical problems of mechanics.\\[0.3cm] 
The closest point projection $\fz = \fP_C\,\fy$ onto $C$, with $\fy\in D$ fixed, follows by minimizing the quadratic function $d^2(\cdot,\fy)$ constrained by
\eqref{eq:constraint1} and \eqref{eq:constraint2}. 
The constraint \eqref{eq:constraint1} is enforced directly by working with the displacement field $\fu\e$ instead of with $\feps\e$. The constraint \eqref{eq:constraint2} is enforced by means of the method of Lagrangian multipliers which will be listed in the vector $\feta$.
Considering $\fy=\{(\feps\es,\fsigma\es)\}_{e=1}^m$ as given from a previous iteration, the corresponding Euler-Lagrange equations can be simplified to the following decoupled linear equation systems
\begin{subequations}
\begin{equation}
\Ass\{w\e\fB\eT\mathbb{C}\e\fB\e\}\fu=\Ass\{w\e\fB\eT\mathbb{C}\e\feps\es\},
\end{equation}
\begin{equation}
\Ass\{w\e\fB\eT\mathbb{C}\e\fB\e\}\feta=\ff-\Ass\{w\e\fB\eT\fsigma\es\}.
\end{equation}
\end{subequations}
The closest state $\fz=\fP_C\,\fy\in C$ then follows by evaluating the new stress and strain for every material point as
\begin{subequations}
\begin{equation}
\feps\e=\fB\e\fu\e, \hspace{1cm} \forall e=1,...,m,
\end{equation}
\begin{equation}
\fsigma\e=\fsigma\es+\mathbb{C}\e\fB\e\feta\e, \hspace{1cm} \forall e=1,...,m.
\end{equation}
\end{subequations}
The strain-stress data of all the material points $e=1,...,m$ are put into $\fz$. 
A very simple data-driven solver then consists of the fixed point iteration 
\begin{equation}
\fz_{j+1}=\fP_C\fP_D\fz_j,\label{eq:fixedppoint}
\end{equation}
according to \citep{kirchdoerfer2016data}
for $j=0,1,...$ and arbitrary $\fz_0\in Z$, where $\fP_D$ denotes the closest point projection in $Z$ onto $D$. 
The fixed point iteration \eqref{eq:fixedppoint} first finds the closest point $\fy_{j+1}=\fP_D\fz_j$ to $\fz_j$ on the material data set $D$ and then projects the result back to the constraint set $C$. 
The iteration is repeated until $\fP_D\,\fz_{j+1}=\fP_D\,\fz_{j}$, i.e., until the assignment of data points in the material data set does not change anymore.\\[0.3cm] 
A special form of the closest point projection $\fP_D$ onto the data set $D$ was introduced as the max-ent data-driven computing paradigm \citep{kirchdoerfer2017data}. 
This formulation generalizes distance-minimizing data-driven computing and is robust concerning noise and outliers. 
Data points are assigned with a variable relevance depending on the distance to the compatible and equilibrated state. 
A max-ent fixed point iteration for a single element is expressed as
\begin{equation}
\fz_{j+1}\e=\fP_C\e\Big(\sum_{i=1}^n p_i(\fz_{j}\e,\beta_{j})\fy_i\Big).
\end{equation}
Here, $\fy_i\in D\e$ are the strain-stress pairs in the material data set and $p_i\in [0,1]$ are the corresponding weights with $\sum_{i=1}^n p_i\e=1$. Further the relation
\begin{subequations}
\begin{equation}
p_i\e(\fz\e_j,\beta_j)=\frac{1}{S(\fz_j\e,\beta_j)}\exp\Big(-\frac{\beta_j}{2}d^2(\fz_j\e,\fy_i)\Big),
\end{equation}
\begin{equation}
S(\fz\e_j,\beta_j)=\sum_{i=1}^n \exp\Big(-\frac{\beta_j}{2}d^2(\fz_j\e,\fy_i)\Big)
\end{equation}
\end{subequations}
holds
with $\beta_j\in(0,+\infty)$ being a Pareto weight which controls the thermalized maximum-entropy extension. 
The annealing schedule in \citep{kirchdoerfer2017data} proposes to update $\beta$ in an annealing schedule of the form
\begin{subequations}
\begin{equation}
\frac{1}{\tilde\beta_{j+1}\e}=\sum_{i=1}^n p_i(\fz_j\e,\beta_j)d^2(\fy_i-\fz\e_{j+1}),
\end{equation}
\begin{equation}
\tilde\beta_{j+1}=\Big(\sum_{e=1}^m w\e\tilde\beta_{j+1}\e\Big)\Big(\sum_{e=1}^m w\e\Big)^{-1},
\end{equation}
\begin{equation}
\beta_{j+1}=\lambda\tilde\beta_{j+1}+(1-\lambda)\beta_k,
\end{equation}
\end{subequations}
where $\beta$ is strictly increasing and finally goes to infinity. 
For $\beta\rightarrow\infty$ the distance-minimizing data-driven scheme is recovered. 
The rapidity of the annealing schedule can be controlled by the parameter $\lambda\in [0,1]$. 
For $\lambda=1$ the fastest evolution of $\beta$ would be achieved, for $\lambda=0$ there is no evolution of $\beta$ at all. 
Note that the initial value $\beta_0$ has to be small enough to ensure the contractivity of the closest point mapping.
\section{Background: Instance-based learning using tensor voting}\label{sec:tensorvoting}
\begin{figure}[htbp]
\centering
 \psfrag{a}{a)}
 \psfrag{b}{b)}
 \psfrag{P1}{$\fP_{\!\!i}$}
 \psfrag{P2}{$\fP_{\!\!j}$}
 \psfrag{v}{$\fv$}
\includegraphics[width=0.6\textwidth]{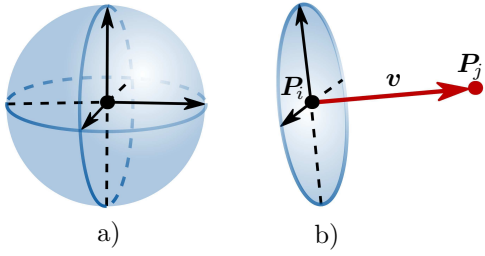}
\caption{Visual representation of a 3D tensor. a) Ball tensor in initial state. b) Tensor after voting.}
\label{fig:ball}
\label{fig:deformation}
\end{figure}

\begin{figure}[htbp]
\centering
 \psfrag{a}{$\sigma=0.1$}
 \psfrag{b}{$\sigma=0.4$}
 \psfrag{c}{$\sigma=1.6$}
\includegraphics[width=\textwidth]{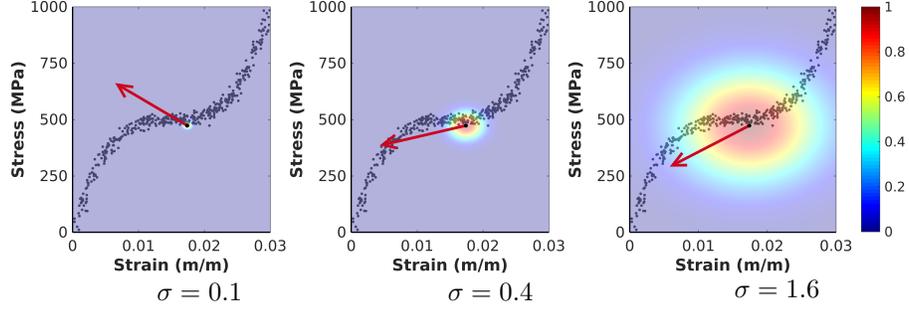}
\caption{Comparison of voting weights $w(\fP_i,\fP_j)$ for three different representations of $\sigma$. 
Colorbar represents values of weights for a point at a certain distance to the voting point $\fP_i$ (black). 
Red arrows head in the direction of corresponding tangents $\ft$.}
\label{fig:voting}
\end{figure}

\noindent
The objective of instance-based learning is to understand the underlying relationships between observations, which are points in an $N$-dimensional continuous space, under the assumption that they lie in a limited part of the space, typically a manifold (see \citep{mitchell1997machine}). 
The principal of learning low dimensional embeddings from points in high dimensional spaces is therefore often referred to as manifold learning.
This field of research was opened by the development of algorithms based on principal component analysis (PCA) by Jolliffe in 1986 \citep{jolliffe1986principal} and multidimensional scaling (MDS) by Cox \& Cox in 1994 \citep{cox1994multidimensional}.
These approaches are based on the idea that nonlinear manifolds can be approximated by locally linear parts.
Schölkopf et al. \citep{scholkopf1997kernel} introduced the kernel PCA, an extension to nonlinear problems, which maps the data implicitly into a higher-dimensional space to extract more information via principal components.
The method of locally linear embedding (LLE) \citep{roweis2000nonlinear} is based on the idea that each point can be reconstructed by its neighbors, if data lies on a locally linear, low-dimensional manifold and the neighbors are associated with appropriate weights.
The method ``isomap'' was presented in Tenenbaum et al. \citep{tenenbaum2000global} and is an extension of MDS. Isomap uses geodesic distances instead of Euclidean distances. 
For this reason, distances between points can be evaluated by graph distances. In this way, the capturing of nonlinearities is facilitated.
Compared with this, the Laplacian eigenmaps algorithm \citep{belkin2003laplacian} can be seen as a generalization of LLE and represents another well-known algorithm in the field of manifold learning.
For detailed information towards manifold learning and further extensions for the methods briefly mentioned above, we refer to \citep{mordohai2010dimensionality}.\\[0.3cm] 
Here, we summarize the method of tensor voting introduced by Mordohai \& Medioni \citep{mordohai2010dimensionality} which is a further development of \citep{mordohai2005unsupervised}, an earlier work of the authors.
Further investigations and extensions are addressed in \citep{moreno2011improving} and \citep{wu2011closed}.
Considering the present application of interest, the method provides the following advantages:
\begin{itemize}
 \item The learning procedure is performed directly in the space of provided data and no mapping to lower or higher dimensions has to be executed.
 \item The method directly provides tangents and normals to the investigated manifold.
 \item Large sets of data can be handled which might contain outliers.
 \item The learning procedure is performed in an offline step.
 \item Data sets with different data distributions and data densities can be handled.
 \item The method is not limited to a certain dimensionality.
 \item The implementation is still sufficiently simple.
\end{itemize}
The tensor voting method is a pairwise operation in which elements cast and collect votes in local neighborhoods.
Each vote is a symmetric, second-order, positive semi-definite tensor and encodes the orientation the receiver would have according to the voter if the voter and receiver belonged to the same structure. 
Tensor voting is based on strictly local computations in the neighborhoods of the inputs. 
In the method presented in \citep{mordohai2010dimensionality} three different cases for computing votes are presented based on the eigenvalues of the initial tensor $\fT_i$ at each point $i=1,...,n$.
These are specified as stick tensors, generic tensors and ball tensors.
For reasons of simplicity, we will assume an unoriented initial tensor, which can be represented, e.g. as a ball (ellipsoid) for the three-dimensional case (see Fig. \ref{fig:ball}a). 
To ensure brevity, the summarizing explanation will be limited to ball tensors only.
A vote for a ball tensor is computed as
\begin{equation}\label{eqn:ballvote}
\fB_{vote}(\fP_i,\fP_j)=\underbrace{e^{-(\frac{s^2}{\sigma^2})}}_{w(\boldsymbol P_i,\boldsymbol P_j)}\big(\fI-\frac{\fv\fv^T}{\norm{\fv^T\fv}}\big),
\end{equation}
where $\fv$ is the vector connecting the receiver $\fP_i\in\mathbb{R}^N$ with the voter $\fP_j\in\mathbb{R}^N$ while $s=\norm{\fv}$ is the corresponding Euclidean distance.
Note, that a voting tensor $\fB_{vote}$ always has one zero-eigenvalue and $N-1$ eigenvalues being equal to the weight $w(\boldsymbol P_i,\boldsymbol P_j)$. 
For the three-dimensional case, the structure can be interpreted as a plane with normal $\fv$ (see Fig. \ref{fig:ball}b). 
 In the depicted example, point $\fP_i$ is the receiver collecting one ball vote by the voter $\fP_j$.
The resulting ball tensor is characterized by two non-zero-eigenvalues and the corresponding eigenvectors which are orthogonal to the voting direction $\fv$.\\[0.3cm] 
The only parameter to be chosen is $\sigma$ which is a control parameter determining the radius of influence \mbox{(see Fig. \ref{fig:voting})}. 
If $\sigma$ is small, a small number of closest points will be taken into consideration for voting. 
If $\sigma$ is chosen to be larger, the range of points with influence on the receiver will be increasing. 
Regarding the choice of $\sigma$, one should remark that $\sigma$ can be converted into an equivalent strain or stress radius.
To determine those radii, a certain weight level has to be specified where votes with lower weights $w$ are not relevant for the solution.
The voting process is completed by the accumulation of all $K$ nearest neighbors' votes
\begin{equation}
\fR_i=\sum_{j=1}^K\fB_{vote}(\fP_i,\fP_j),
\end{equation}
where the sum can simply be computed by adding the corresponding $N\times N$-matrices.\\ [0.3cm] 
Followed by the voting process, the accumulated votes have to be interpreted in an analyzing step. 
By means of spectral decomposition and analyzing the ``received'' (i.e.\ the computed) eigenvalues the dimensionality $k$ of the underlying data structure can be estimated.
Further, the computation of eigenvectors is performed in order to receive tangents and normals to the manifold, which are the relevant outputs for the following data-driven computations. 
The eigenvectors with the $N-k$ highest eigenvalues are denoted as normals and the eigenvectors with the $k$ lowest eigenvalues will be used as tangents.
The eigenvectors are then ordered in a matrix $\fT_i$ according to descending eigenvalues:
\begin{equation}
 \fT_i=\sv | & & | & | &  & | \\ \fn^1_i & ... & \fn^{N-k}_i & \ft^{1}_i & ... & \ft^k_i \\ | &  & | & | & & |  \ev = \sv \fT^n_i & \fT^t_i \ev,
\end{equation}
with $\fT^n_i=\{\fn^l_i\}_{l=1}^{N-k}$ being the normals and $\fT^t_i=\{\ft^l_i\}_{l=1}^{k}$ being the tangents to the manifold at point $i$. 
Fig. \ref{fig:voting} shows the tangents as voting results for three different choices of $\sigma$ with the corresponding relevance to the voting procedure. 
It should be emphasized that in the context of material data consisting of strain and stress pairs the previously introduced distance metric \eqref{eq:distmetric} is used.
To deploy the Euclidean distance in the voting process, a point for voting is represented as
\begin{equation}\label{equ:learningspace}
 \fP_i = \sv \mathbb{C}^{1/2}\feps_i \\ \mathbb{C}^{-1/2}\fsigma_i \ev,
\end{equation}
where based on the symmetric and positive definite matrix $\mathbb{C}$ introduced in \eqref{eq:distmetric} stress and strain components are mapped into an intermediate state, which we will denote as the learning space.
Then the squared distance $s^2$ between two points used for voting is twice the distance computed from the squared metric \eqref{eq:distmetric}.
A simple procedure for tensor voting using ball tensors only is shown in Algorithm \ref{alg:tensorVoting}.
\begin{algorithm}[H]
\caption{Tensor voting}
\label{alg:tensorVoting}
\begin{algorithmic}
\REQUIRE data set $D$, control parameter $\sigma$ and number of nearest neighbors $K$
\FORALL{$i=1,\dots,n$}
\STATE Initialize $\fR_i$ to zero.
\FORALL{$j=1,\dots,K$}
\STATE
\begin{equation}
\fv=\fP_i-\fP_j
\end{equation}
\STATE
\begin{equation}
s=\norm{\fv}
\end{equation}
\STATE
\begin{equation}
  \fB_{vote}=e^{-(\frac{s^2}{\sigma^2})}\big(\fI-\frac{\fv\fv^T}{\norm{\fv^T\fv}}\big).
\end{equation}
\STATE
\begin{equation}
  \fR_i=\fR_i+\fB_{vote}
\end{equation}
\ENDFOR
\STATE Compute spectral decomposition of $\fR_i$ and order the eigenvectors in matrix $\fT_i$ according to descending eigenvalues.
\ENDFOR
\end{algorithmic}
\end{algorithm}

\section{Extension of data-driven computing using pointwise tangent spaces} \label{sec:ddtensorvoting}
\begin{figure}[htbp]
\centering
\includegraphics[width=0.45\textwidth]{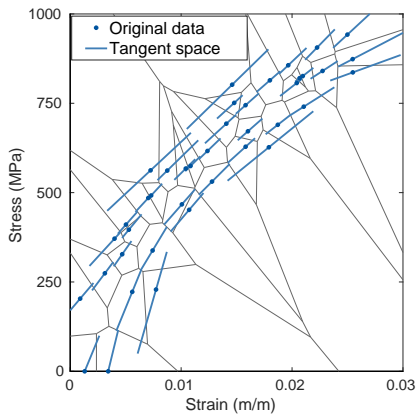}
\caption{Voronoi tessellation of a data set with corresponding tangent spaces.}
\label{fig:voronoi}
\end{figure}

\noindent
As stated in Chapter \ref{sec:datadriven} the data-driven framework consists of two projections $\fP_C$ and $\fP_D$. 
The former is to satisfy kinematic and equilibrium conditions, whereas the latter is to find new states $\{\fy\e\}_{e=1}^m$ for stress and strain in local material data sets $D\e$. 
In this work we extend the closest point projection $\fP_D$ by means of a new projection $\fP_T$ onto pointwise tangent spaces $\{\fT_i\}_{i=1}^n$. 
For the generalized case of entropy-maximizing data-driven computing the adapted fixed point iteration follows as
\begin{equation}
\fz_{j+1}\e=\fP_C(\sum_{i=1}^n p_i(\fz\e_j,\beta_j),\fx_i\e)
\end{equation}
with
\begin{equation}
\fx_i\e=\fy_i\e+\fT^t_i\boldsymbol\lambda^t_i
\end{equation}
or more generally
\begin{equation}
 \fx_j\e=\fP_T\e(\fz\e_j),
\end{equation}
where $\fy_i\e\in D\e$ denotes the closest point to the state $\fz\e_j$ at iteration $j$. 
In contrast to the initial formulation, $\fx_i\e\in\mathbb{R}^{2m_e}$ is not a member in the material data set but the closest point to the data set extended by pointwise tangent spaces.
The quantities $\boldsymbol\lambda_i^t,\,\boldsymbol\lambda_i^n\in\mathbb{R}^{m_e}$ is computed by an orthogonal projection $\fP_T\e$ of $\fz\e_j$ onto the manifold, satisfying the condition $\fz\e_j-\fy_i\e=\fT_i\boldsymbol\lambda_i$ with $\boldsymbol\lambda_i=(\boldsymbol\lambda_i^n \;\;\boldsymbol\lambda_i^t)^T\in\mathbb{R}^{2m_e}$.
Global iterations are performed until $\beta$ exceeds the predefined value $\beta_{End}$.\\[0.3cm]
In the limit of $\beta \rightarrow \infty $ the distance-minimizing scheme is recovered.
Due to the extended data set with
\begin{subequations}
\begin{align}
 \feps\es&=\bar\feps^{(e)}+\bar\fT^t_\varepsilon\bar{\boldsymbol\lambda^t},\\
 \fsigma\es&=\bar\fsigma^{(e)}+\bar\fT^t_\sigma\bar{\boldsymbol\lambda^t},
\end{align}
\end{subequations}
the formulation simplifies the location of the closest point $\fx\e_{j}=(\feps\es,\fsigma\es)$.
In the latter equation, $\fy\e_{j}=(\bar\feps^{(e)},\bar\fsigma^{(e)})$ denotes the closest point to local state $\fz_j\e$ in the material data set $D\e$ and $\bar\fT$ is the corresponding tangent matrix, which can be split into a stress $(\bullet)_\sigma$ and strain $(\bullet)_\varepsilon$ related part
\begin{equation}
  \bar\fT= \sv \bar\fT^n_\varepsilon & \bar\fT^t_\varepsilon \\ \bar\fT^n_\sigma & \bar\fT^t_\sigma \ev,
\end{equation}
with $\bar\fT^n_\varepsilon$, $\bar\fT^t_\varepsilon$, $\bar\fT^n_\sigma$, $\bar\fT^t_\sigma\in\mathbb{R}^{m_e\times m_e}$.
Here, global iterations are performed until the distance $d(\fz_{j+1},\fx_{j+1})$ is not lower than in the iteration before or when a certain tolerance is reached.\\[0.3cm]
The presented method enables to remain as close as possible to the given data. 
For this purpose, first of all, we search for the point in the material data set which is closest to the actual state.
Secondly, taking this information into account, we look for the nearest point in the corresponding tangent space. 
The distribution of points, which can be theoretically reached is depicted in Fig. \ref{fig:voronoi}. 
Let us consider the search procedure in the given pictogram for a given compatible and equilibrated state. 
First, the closest material state in the data set is associated in accordance with the Voronoi voxel 
in which the current local state is positioned.
Second, the mapping onto the tangent space is performed. 
It can be observed that the points found in the tangent space do not necessarily have to be within the same Voronoi voxel. 
Another remark concerns the structure of the tangent spaces. 
As we are dealing with tangent and normal vectors spanning a linear manifold the calculation of the closest point projection onto the manifold is straightforward.
The projection can be done by simply solving a linear equation system of dimension $2m_e$ to compute a unique solution. 
All matrices $\fT_i$ can be decomposed during the offline step to improve efficiency significantly.
\section{Numerical results for trusses}\label{sec:trusses}
\begin{figure}[htbp]
\begin{subfigure}{0.59\textwidth}
\centering
\includegraphics[width=1.15\textwidth]{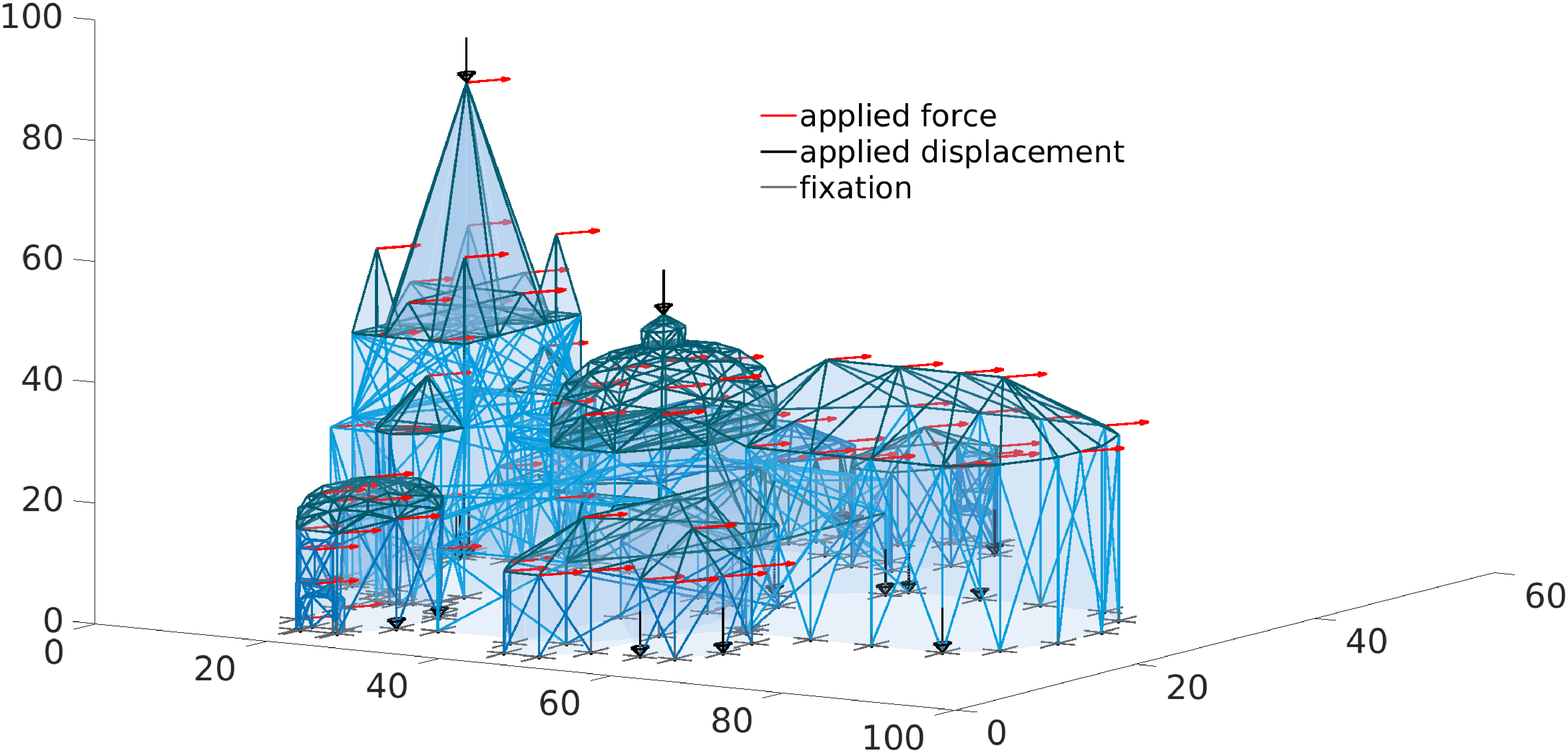}
\subcaption{}
\label{fig:truss}
\end{subfigure}
\begin{subfigure}{0.39\textwidth}
\centering
\includegraphics[width=\textwidth]{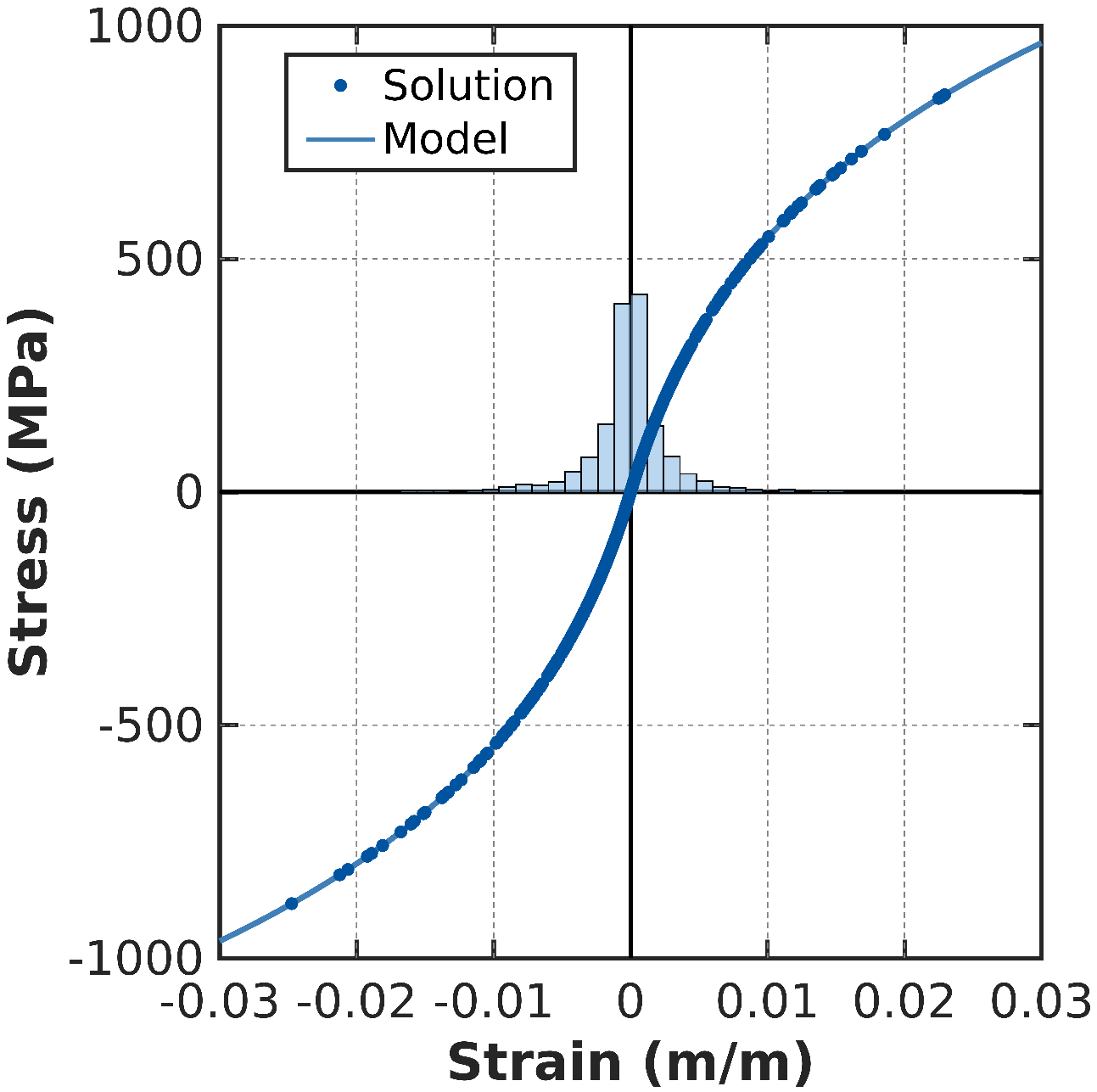}
\subcaption{}
\label{fig:nonlinearlaw}
\end{subfigure}
\caption{a) Truss geometry (inspired by the Aachen Cathedral), and boundary conditions. 
Red arrows represent applied loads, black arrows prescribed displacements. 
Nodes with grey crosses are fixed. 
b) Reference material model and solution points of the truss problem. 
Histogram visualizing the occurrence of strain solutions.}
\end{figure}

\noindent
We demonstrate the convergence properties of the tensor voting data-driven extension by the aid of a three-dimensional truss structure.
The geometry of the truss, which comprises $m=1,513$ bars, the boundary conditions and the applied loads are shown in Fig. \ref{fig:truss}. 
In total the problem under consideration comprises 1,234 degrees of freedom. 
The truss undergoes small deformations and the material in all bars obeys the nonlinear elastic material law shown in Fig. \ref{fig:nonlinearlaw}. 
The reference solution, which is also plotted in Fig.\ \ref{fig:nonlinearlaw} by points $\fz\e_{ref}=(\varepsilon\e_{ref},\sigma\e_{ref}$) for all $e=1,...,m$ on the reference stress-strain curve, was computed by a Newton-Raphson solver.

\subsection{Noise free material data}
\begin{figure}[htbp]
\centering
\begin{subfigure}{0.49\textwidth}
\centering
\includegraphics[width=0.97\textwidth]{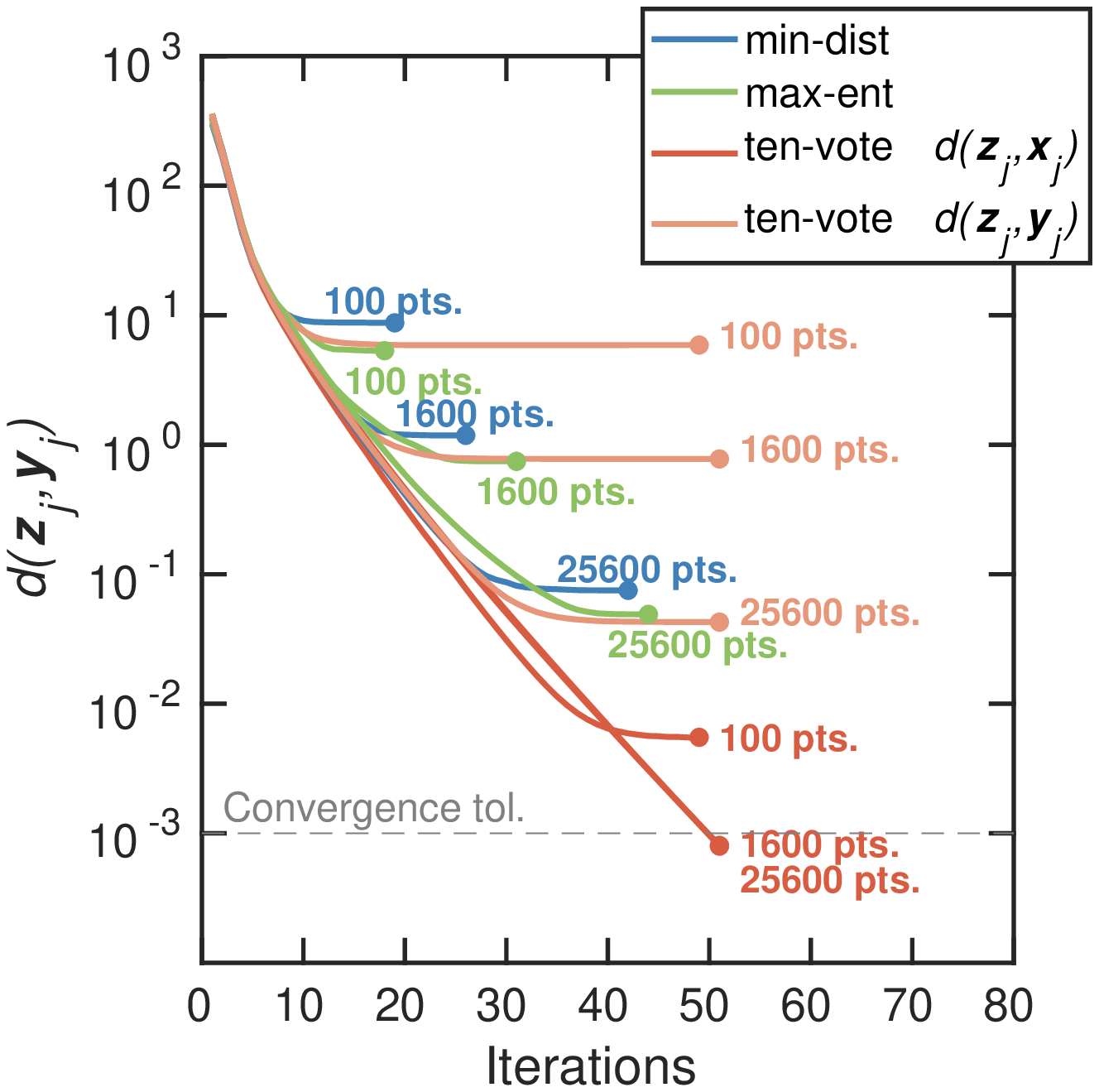}
\subcaption{}
\label{fig:dataassign}
\end{subfigure}
\begin{subfigure}{0.49\textwidth}
\centering
\includegraphics[width=0.97\textwidth]{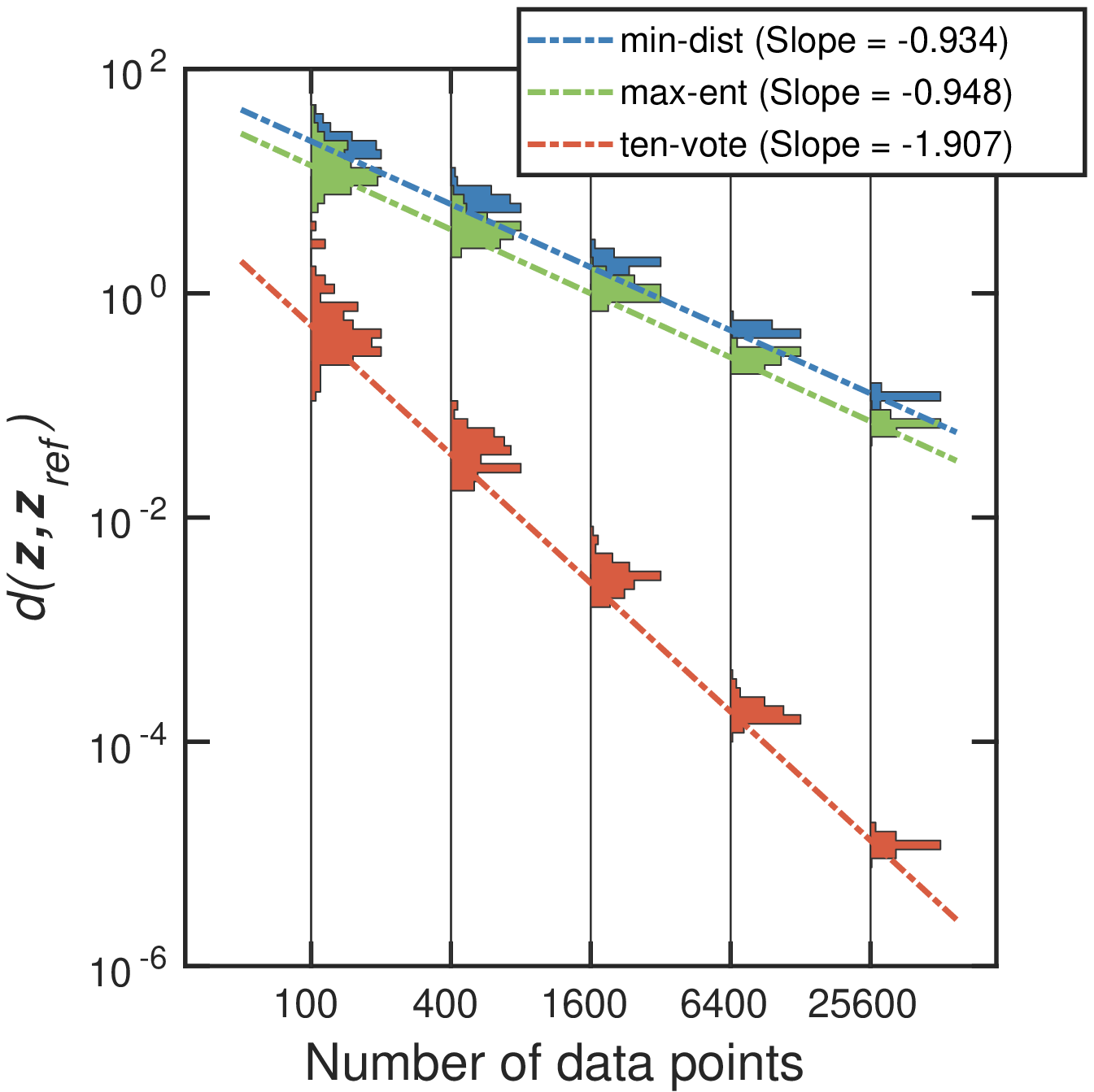}
\subcaption{}
\label{fig:convhistoideal}
\end{subfigure}
\caption{Truss test case. a) Convergence of the local data assignment iteration of the min-dist, max-ent and tensor voting data-driven solution for a data set free of noise. Distance to actual data points and to tangent space plotted for the tensor voting solution. b) Comparison of convergence with respect to data set size of various data-driven methods for a noise-free data set with randomly distributed data points. Error histograms generated from 100 material set samples.}
\end{figure}

\begin{figure}[htbp]
\centering
\psfrag{Number of data points}{Number of data points}
\includegraphics[width=0.5\textwidth]{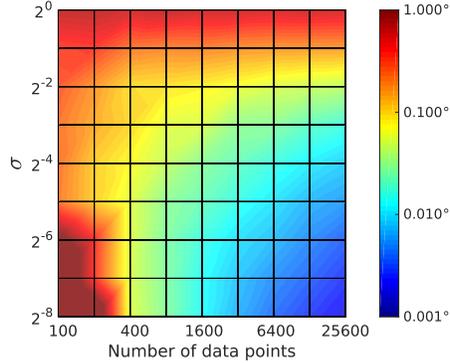}
\caption{Voting performance for noise-free data sets for various data set sizes and training parameters $\sigma$. Average angular error between learned tangent and tangent computed from the reference function in degree.}
\label{fig:tensorVoteIdeal}
\end{figure}

\noindent
We begin by considering a sequence $(D_h)$ with $h=1,...,5$ increasingly fine data sets consisting of $25\cdot4^h$ points on the stress-strain curve with a random distribution of $\varepsilon\in[-0.025,\; 0.025]$.
The convergence of the local data assignment iteration is shown in \mbox{Fig. \ref{fig:dataassign}} for data sets $h=1,3,5$. 
In all cases, the initial local data assignment is random and convergence is monitored in terms of the distance \eqref{eq:norm}. 
The shown plots for the max-ent solver were computed with $\lambda=0.5$.
The solutions of the min-dist and ten-vote solvers were computed with $\sigma_h=(1/4)^h$, where $\sigma$ was introduced as a parameter in the ball voting algorithm (see Eqn. \ref{eqn:ballvote}).
The size of $\sigma$ is decreasing with the expected average distance between the data points.\\[0.3cm]
\mbox{Fig. \ref{fig:dataassign}} shows the evolution of distances for the min-dist, max-ent, as well as the corresponding tensor voting algorithms.
In addition, regarding the tensor voting solver, two different distances are plotted, namely the distance to the actual data points $d(\fz_j,\fy_j)$ and the distance to the tangent spaces $d(\fz_j,\fx_j)$.\\[0.3cm]
The final distance of the min-dist algorithm is outperformed by the max-ent solution for all data samples.
This comes as expected, since the max-ent solver is supposed to find the total minimum of the problem as shown in \citep{kirchdoerfer2017data}.
Remarkably, the number of iterations the max-ent solver needs to find the solution for the here investigated example is comparable to the number of iterations performed for the min-dist solver.
Considering the solution of the ten-vote solver, it is observed that the final distances to the actual data set are very close to the ones of the max-ent solver.
The distance to the tangent spaces is decreasing for all sampling sizes below the minimum distance of the finest sampling computed with the max-ent solver.
For the two finer samples, the convergence tolerance is achieved before the curve for the distance $d\,(\fz_j,fy_j)$ reaches a horizontal tangent.\\[0.3cm] 
Now, we turn to the question of convergence towards the reference solution $\fz_{ref}=\{\fz_{ref}\e\}_{e=1}^m$ with respect to the number of data points.
For additional information, we monitor the convergence of the resulting sequence of data-driven solutions to the reference solution in the sense of distance $d(\fz,\fz_{ref})$ depicted in \mbox{Fig. \ref{fig:convhistoideal}}.
As observed from the diagram, the convergence of the min-dist and the max-ent data-driven solver is close to linear. 
Using the max-ent data-driven solver the rate remains nearly unchanged, but the total error is smaller.\\[0.3cm]
The performance of the here presented tensor voting extension to the data-driven framework known from earlier publications is highly satisfactory. 
We observe a convergence rate which is approximately quadratic. 
Therefore, we can interpret the method as a higher- or second-order data-driven scheme. 
In addition, the total error is more than an amplitude smaller, even for the smallest sampling size. 
It should be remarked that the distance to the classical data-driven solution depends on the nonlinearity and smoothness of the constitutive behavior. \\[0.3cm]
Next, it is worthwhile to consider the performance of the voting procedure.
The tensor voting algorithm is applied to data sets of varying sample sizes and a selection of different values of $\sigma$.
The voting error in dependence of the respective data set size and $\sigma$ is depicted in \mbox{Fig. \ref{fig:tensorVoteIdeal}}.
To quantify the quality of the voting procedure, the average angle of all data samples between the learned tangents and the tangents of the reference function is computed as follows:
\begin{equation}
    \Delta \theta = \frac{1}{N}\sum_{i=1}^N \arccos{\Big(\frac{\ft_i\cdot \ft_i^{Ref}}{|\ft_i||\ft_i^{Ref}|}\Big)}.
\end{equation}
The angular error is computed in the learning space of the data to avoid scaling inconsistencies (see Eq. \ref{equ:learningspace}).
In the case of noise-free data, the performance of the voting procedure can be improved by choosing smaller values for $\sigma$.
The smaller $\sigma$, the fewer points are relevant for the voting procedure.
In this case, the angular error will be the lowest if only the direct neighbors are taken into consideration.
Additionally, with increasing sample sizes, the average distances between the points decrease.
Thus, the error is reducing in the same manner.
\subsection{Maximum entropy approach for noisy material data}
\begin{figure}[htbp]
\centering
\includegraphics[width=0.5\textwidth]{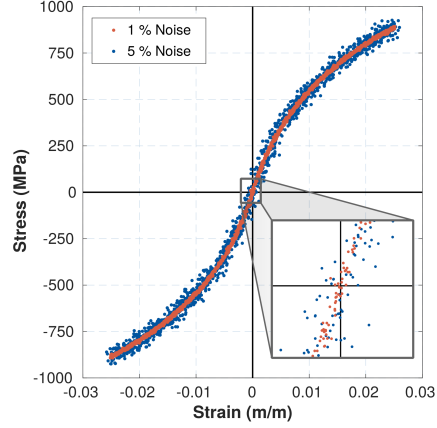}
\caption{Random data sets generated according to normal distribution centered on the material curve of Fig. \ref{fig:nonlinearlaw} with two settings of standard deviation.}
\label{fig:datadistribution}
\end{figure}

\begin{figure}[htbp]
\begin{subfigure}{0.49\textwidth}
\centering
\includegraphics[width=0.97\textwidth]{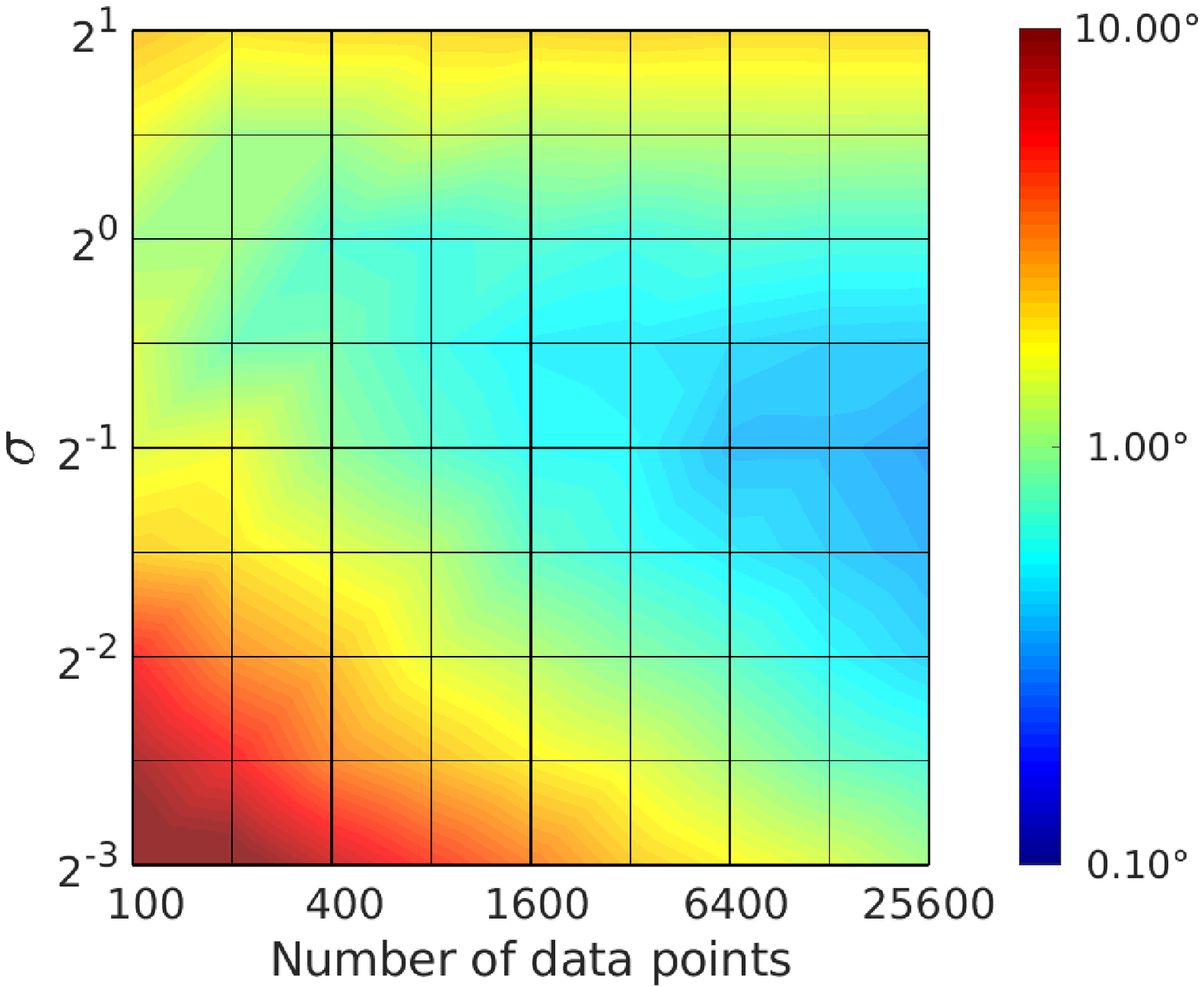}
\subcaption{}
\label{fig:tensorVoteNoise1}
\end{subfigure}
\begin{subfigure}{0.49\textwidth}
\centering
\includegraphics[width=0.97\textwidth]{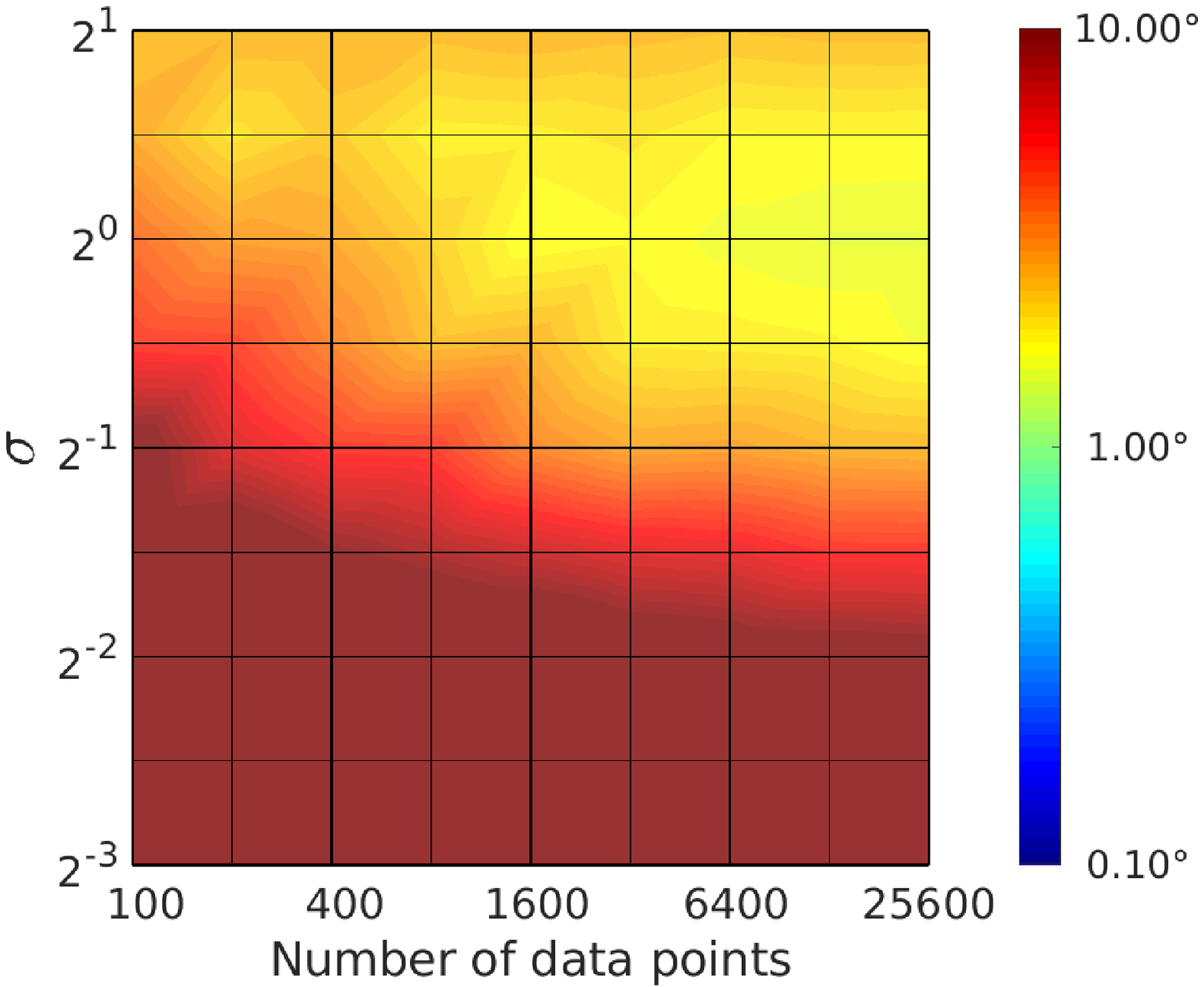}
\subcaption{}
\label{fig:tensorVoteNoise5}
\end{subfigure}
\caption{Voting performance for a) 1\% noise data sets b) 5\% noise data sets for various data set sizes and training parameters $\sigma$. 
Averaged angular error between learned tangents and tangents computed from the closest point on the reference function in degree.}
\end{figure}

\begin{figure}[htbp]
\begin{subfigure}{0.49\textwidth}
\centering
\includegraphics[width=0.97\textwidth]{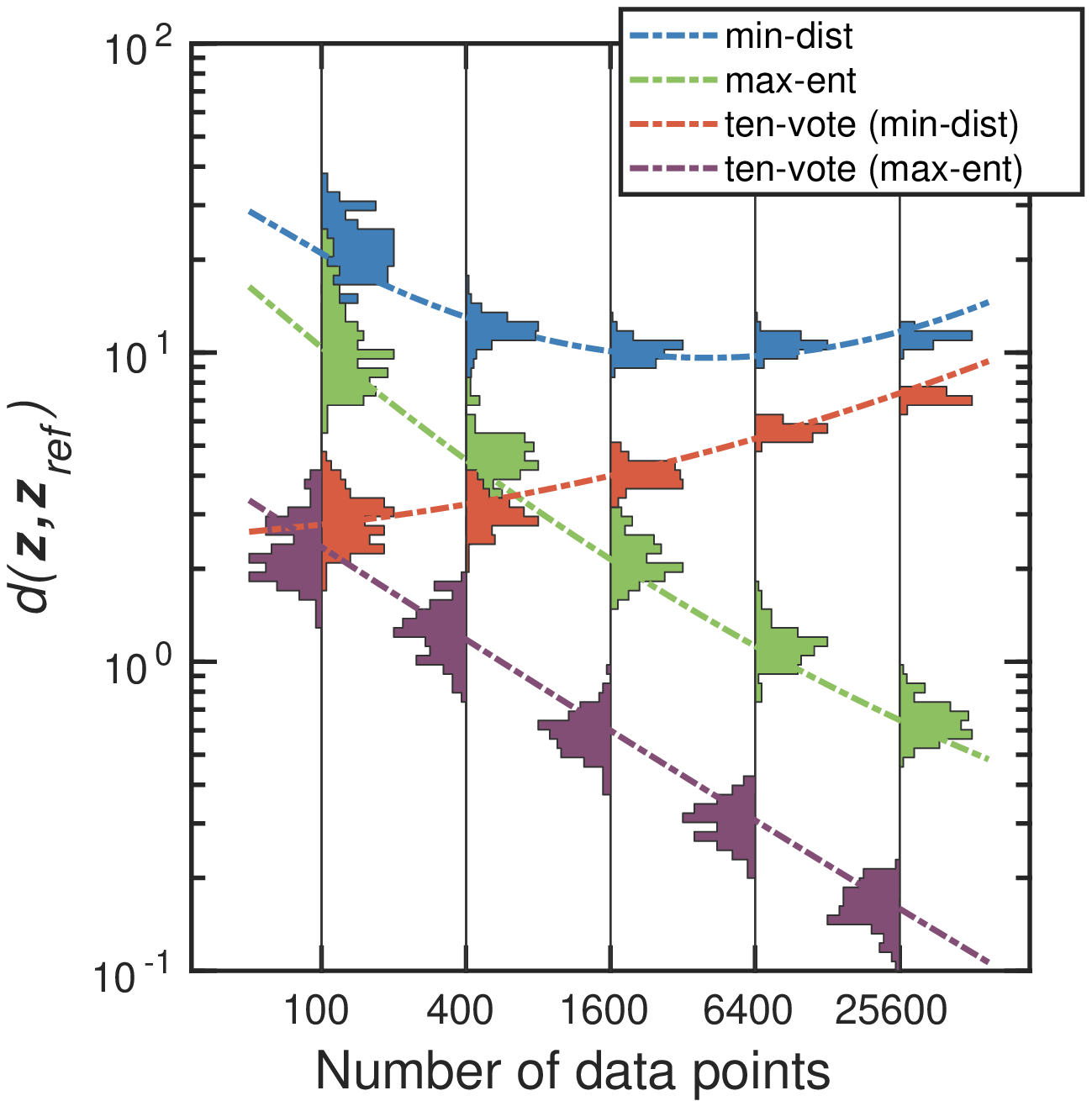}
\subcaption{}
\label{fig:convhistonoise1}
\end{subfigure}
\begin{subfigure}{0.49\textwidth}
\centering
\includegraphics[width=0.97\textwidth]{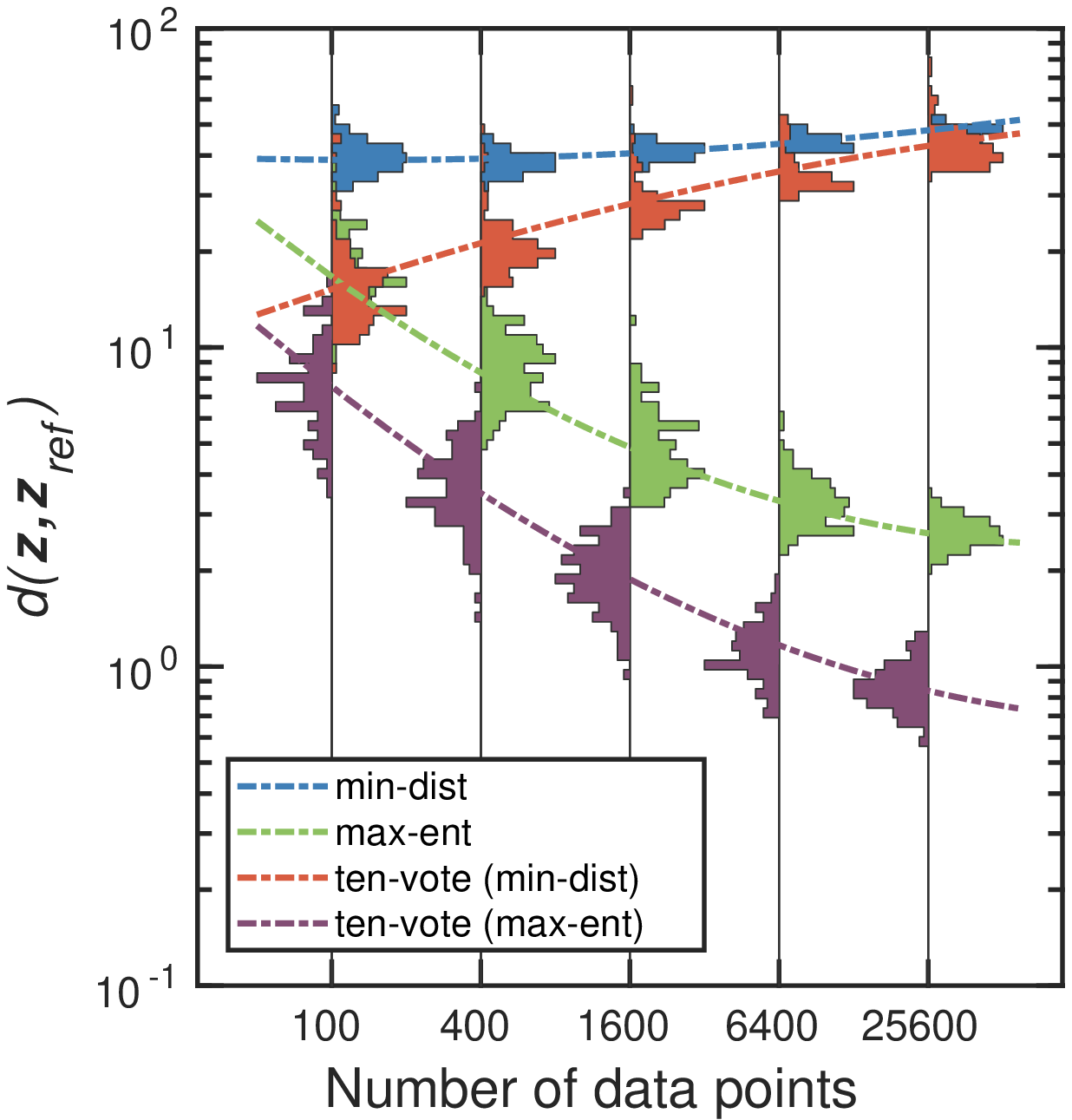}
\subcaption{}
\label{fig:convhistonoise5}
\end{subfigure}
\caption{Comparison of convergence with respect to data set size of various data-driven methods for a) 1\% noise data set b) 5\% noise data set with randomly distributed data points. Error histograms generated from 100 material set samples.}
\end{figure}

\noindent 
Next, we consider data sets that include noise.
The here discussed noise is added by a zero-mean normal distribution with constant standard deviation in strain and stress direction.
Two different settings are investigated and shown in Fig. \ref{fig:datadistribution} for sampling sizes of 1600 points.
The different settings have a standard deviation of \mbox{1\%} and \mbox{5\%} of the maximum values of strains and stresses.
First, the performance of the tensor voting procedure is investigated as performed in the section before.
For different data set sizes and control parameters $\sigma$, the angular error is computed.
The results for the two investigated parameters are depicted in Fig. \ref{fig:tensorVoteNoise1} and Fig. \ref{fig:tensorVoteNoise5}.
This time we consider the angle between the learned tangent and the tangent of the reference function at the position which is closest to the actual data point on the reference function.
In contrast to the section before the best choice for $\sigma$ is not achieved for very small values that give relevance to only a little number of nearest neighbors.
Instead, the dependence of the optimal choice of $\sigma$ on the data set size is decreasing with increasing noise in the data.
In general, a larger noise standard deviation should lead to a higher optimal choice of $\sigma$, since more points are relevant during voting.
For the 1\% noise data set, the optimal choice would be approximately $\sigma=2^{-1}$ and for the 5\% noise data set $\sigma=2^{1}$, respectively.\\[0.3cm]
We revisit the question of convergence towards the reference solution with respect to the data set size of the truss problem.
The convergence plot for 100 data set samplings with \mbox{1 \%} noise is depicted in Fig. \ref{fig:convhistonoise1} and  for samplings with \mbox{5 \%} noise in \mbox{Fig. \ref{fig:convhistonoise5}}, respectively.
All max-ent computations are performed with an annealing control of $\lambda=0.1$.
The computations using the tensor voting solution scheme are done with a learning parameter of $\sigma=0.5$ and a $\beta_{End}=100$ for the 1 \% noise data set and with $\sigma=1.0$ and $\beta_{End}=10$ for the 5 \% noise data set.\\[0.3cm]
In general, the min-dist solution scheme faces the problem that outliers are more likely to be chosen.
In \mbox{Fig. \ref{fig:convhistonoise5}}, this behavior can be clearly identified for the classical min-dist formulation as well as for the tensor voting scheme since the distance to the reference solution is increasing with larger data set sizes.
It is worthwhile to mention that the min-dist method and the ten-vote method perform very well for smaller data set sizes.
The error approximately coincides with the one of the max-ent solutions for the smallest data sampling.
Regarding the max-ent tensor voting solution, the classical max-ent solution scheme is slightly outperformed.
Considering the max-ent tensor voting solution scheme in Fig. \ref{fig:convhistonoise1}, the performance for the 1\% noise data set  is remarkably good, especially for small data set sizes.
For the coarsest data set size, the error more or less coincides with the solution of the min-dist tensor voting solution scheme.
Summarizing, the newly proposed tensor voting method always gives the best results when combined with the max-ent regularization.
\section{Numerical results for 2D problem} \label{sec:continuum}
\begin{figure}[htbp]
\begin{subfigure}{0.49\textwidth}
\centering
 \psfrag{m1}{$100\,mm$}
 \psfrag{m2}[0][0][1][180]{$100\,mm$}
 \psfrag{m3}{$25\,mm$}
\includegraphics[height=6cm]{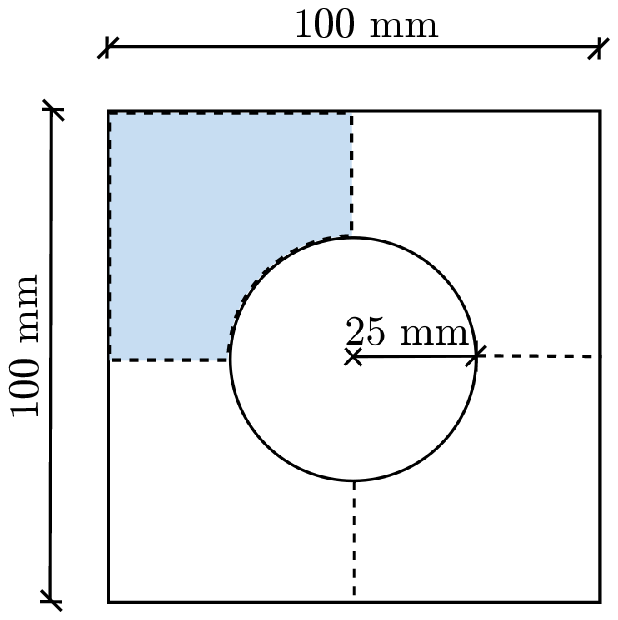}
\caption{}
\label{fig:plateWithHole}
\end{subfigure}
\begin{subfigure}{0.49\textwidth}
\centering
 \psfrag{F}{$q(x)$}
 \psfrag{x}{$x$}
 \psfrag{y}{$y$}
\includegraphics[height=6cm]{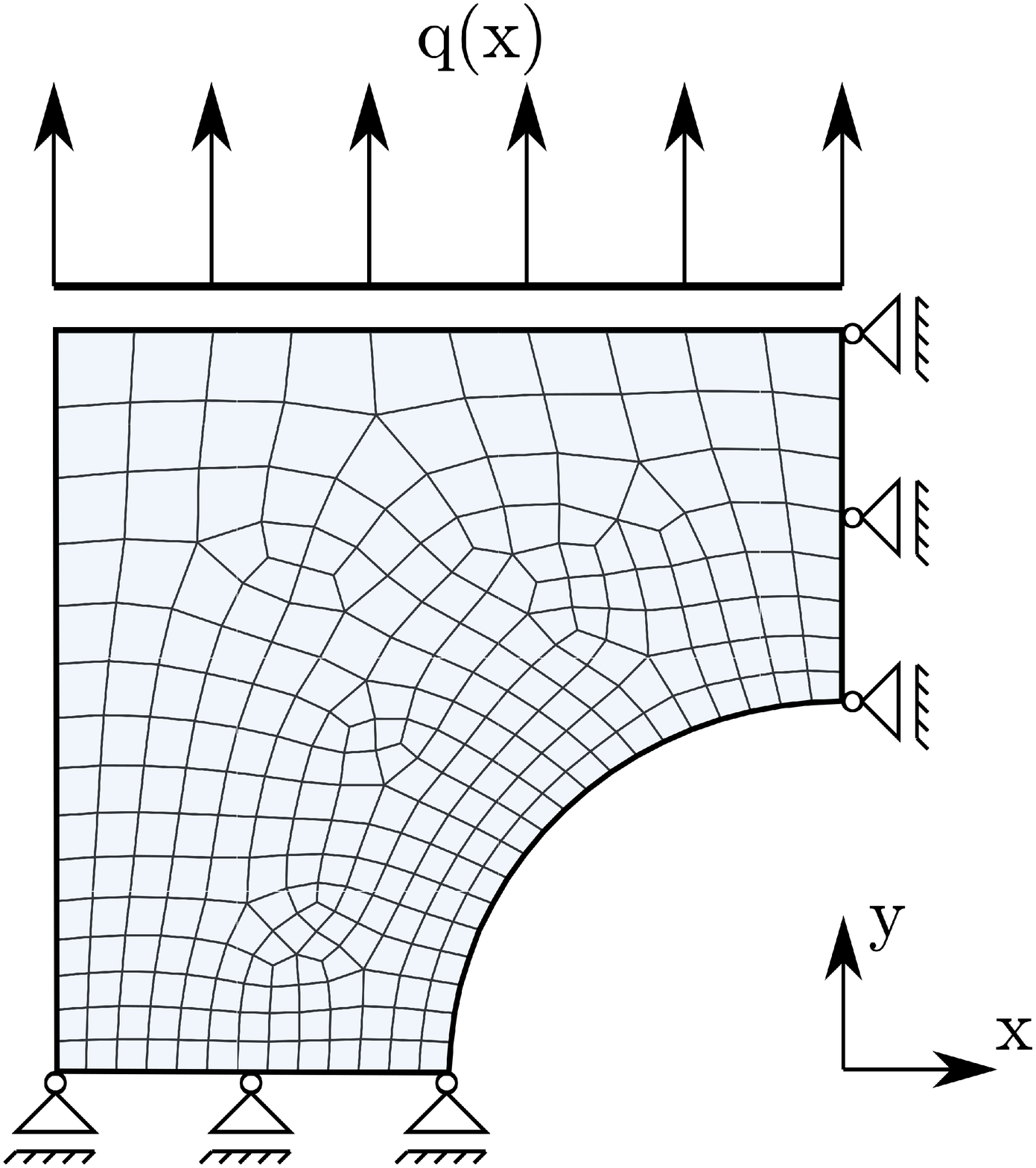}
\caption{} 
\label{fig:boundaryPWH}
\end{subfigure}
\caption{a) Geometry of a plate with a hole. b) Quarter of the system with boundary conditions and mesh discretization.} 
\end{figure}

\begin{figure}[htbp]
\centering
\includegraphics[width=0.6\textwidth]{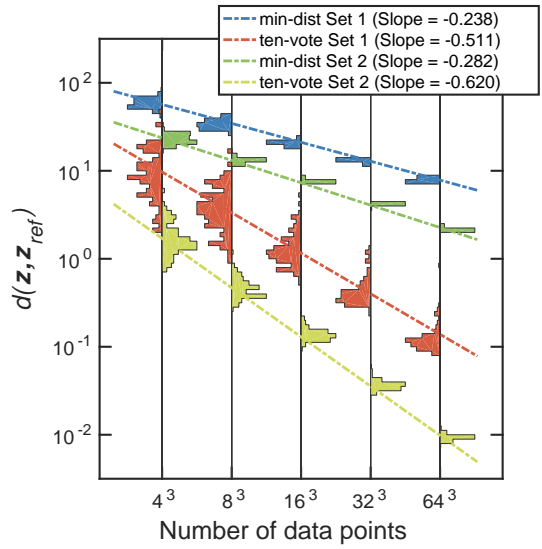}
\caption{Comparison of convergence with respect to data set size of min-dist data-driven methods for noise-free data sets with random but equally distributed (\mbox{set 1}) and random normally distributed (\mbox{set 2}) data points. Error histograms generated from 100 material set samples.}
\label{fig:convhistoideal2D}
\end{figure}

\begin{figure}[htbp]
\centering
\includegraphics[width=0.49\textwidth]{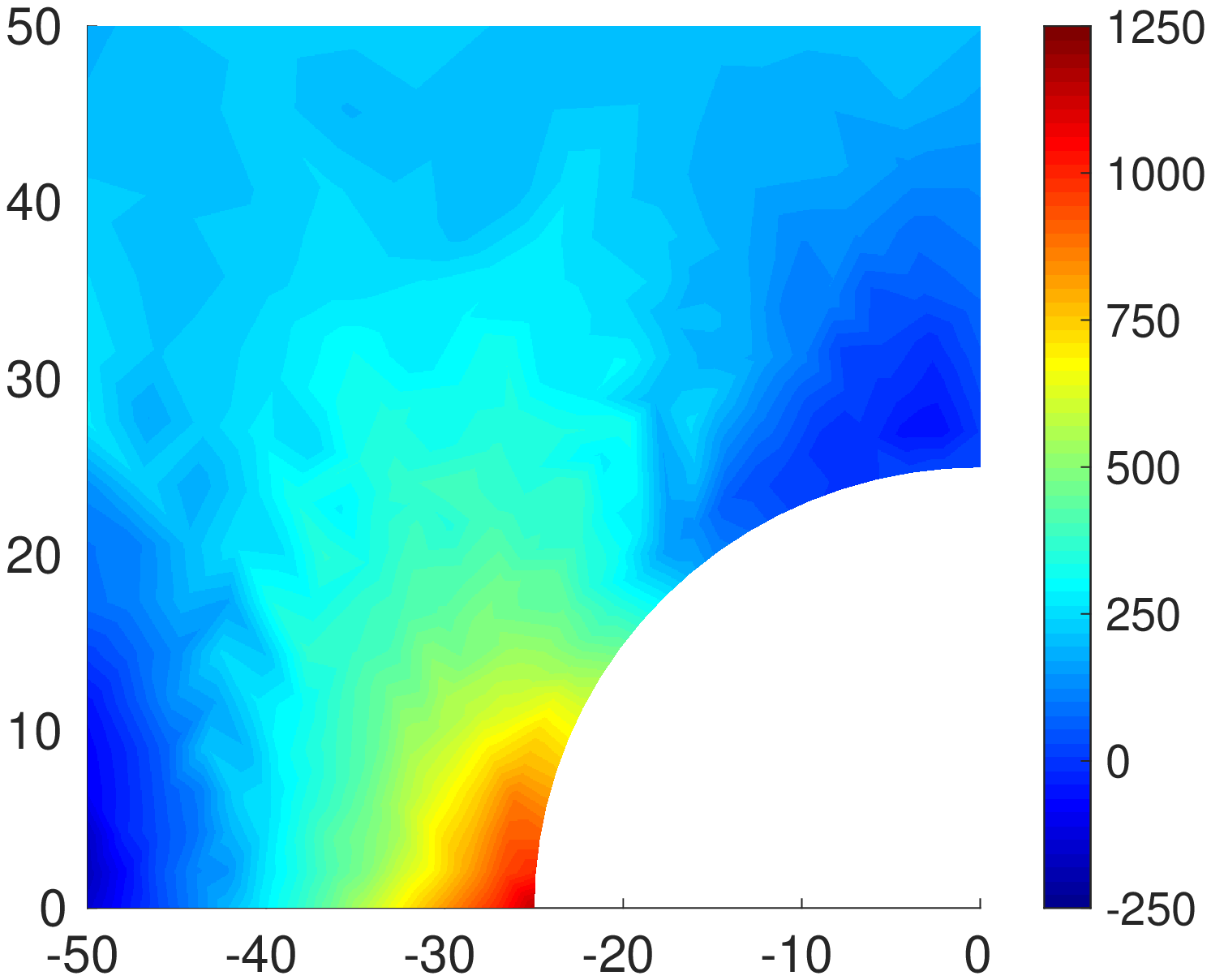}
\includegraphics[width=0.49\textwidth]{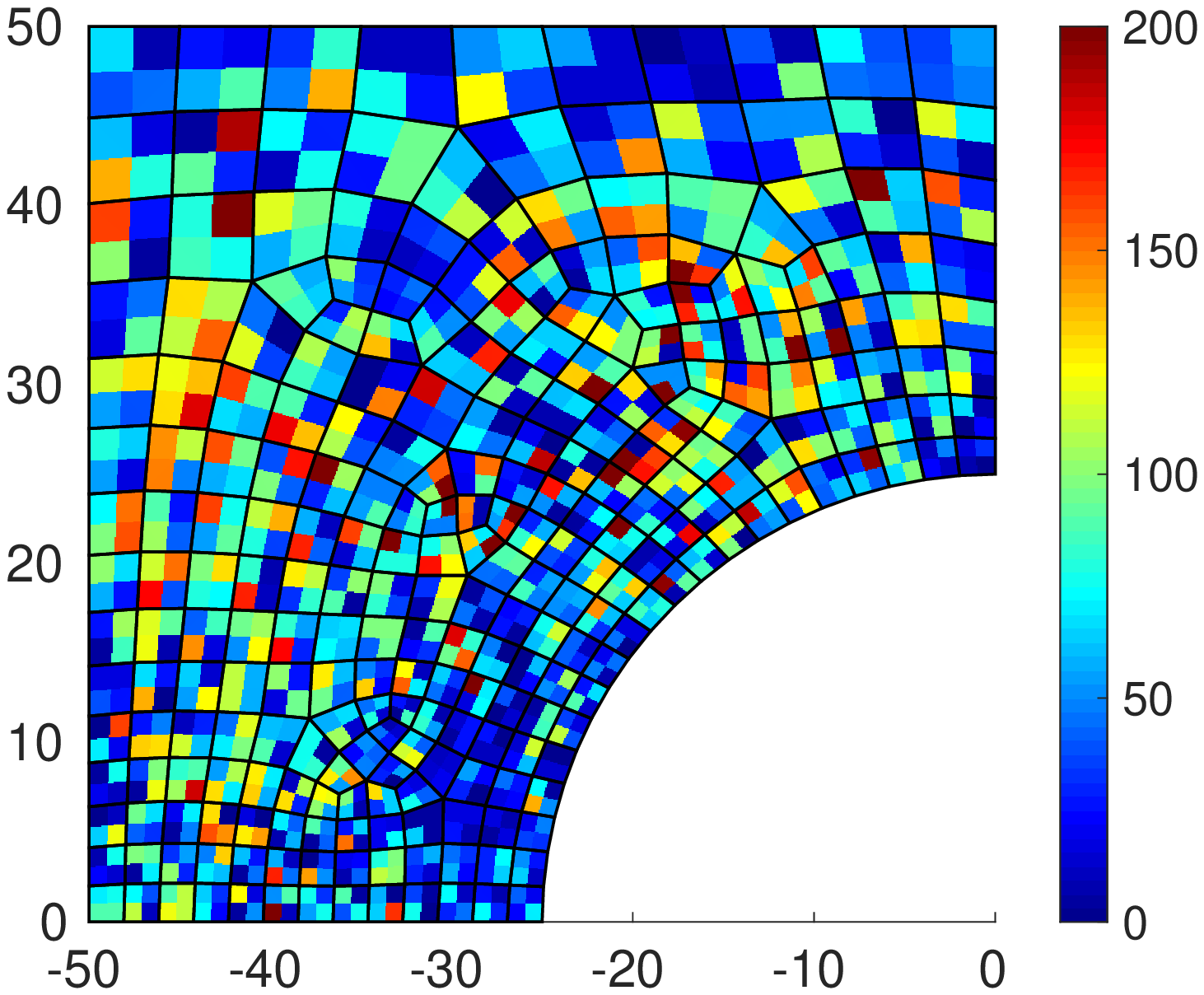}
\includegraphics[width=0.49\textwidth]{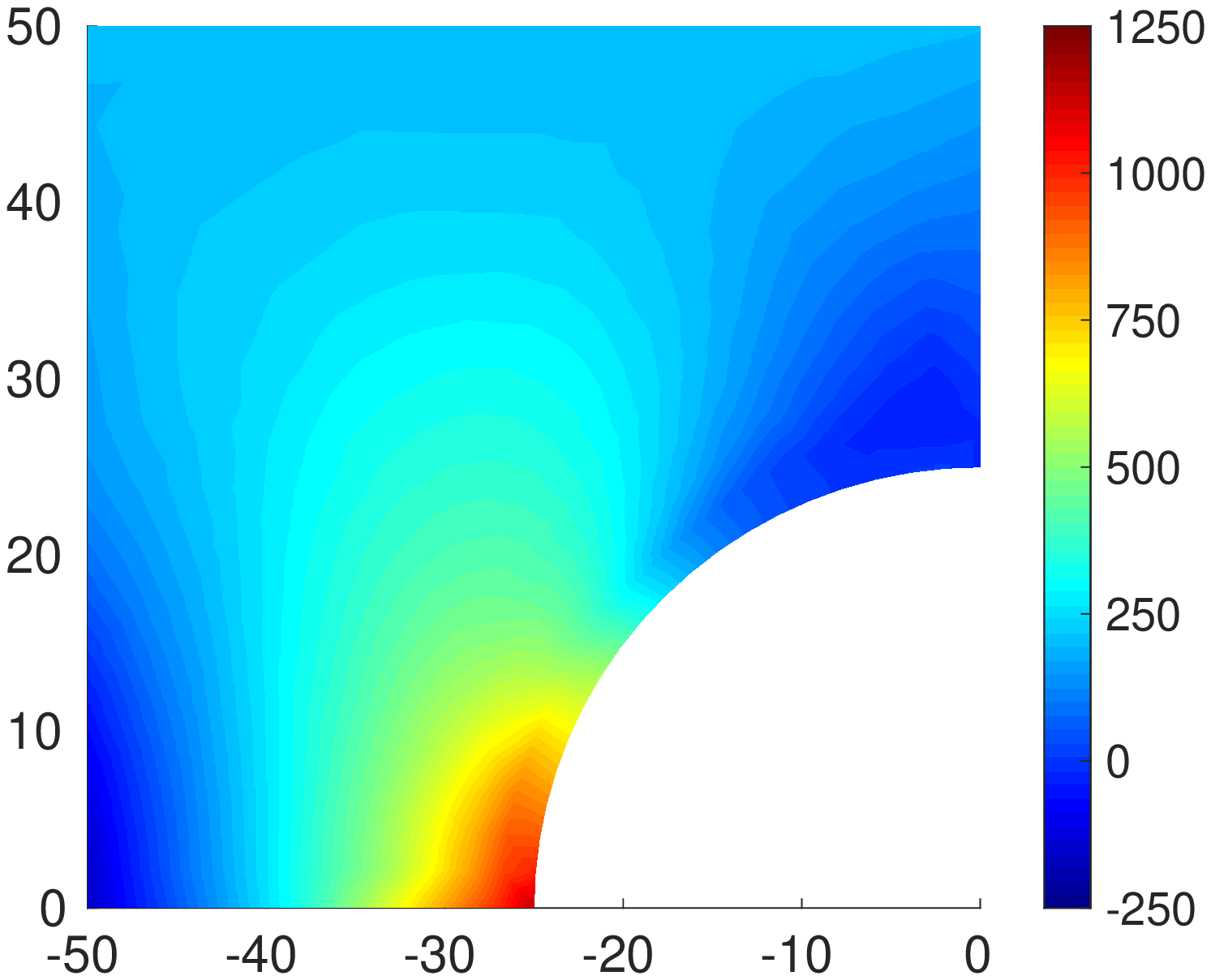}
\includegraphics[width=0.49\textwidth]{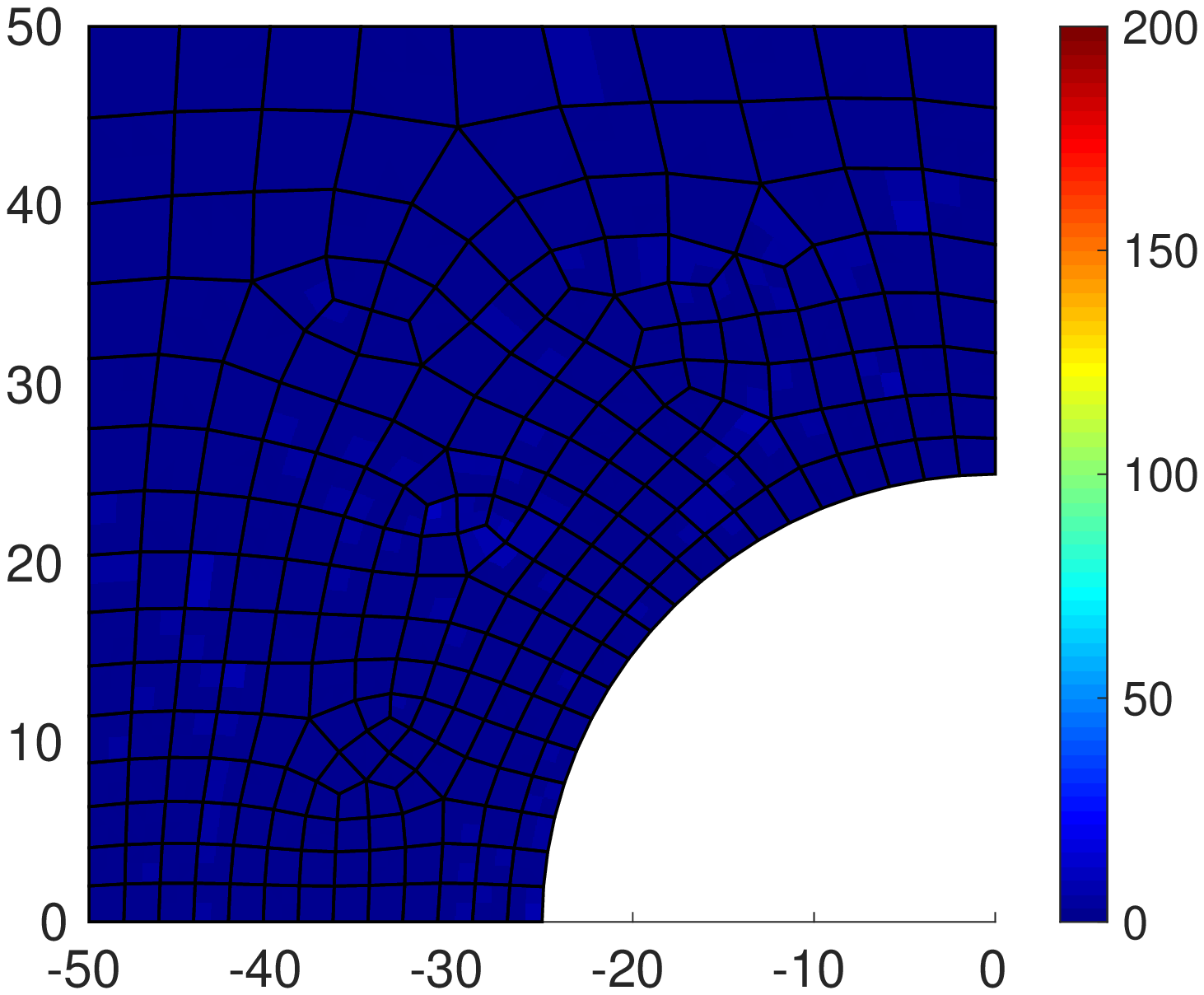}
\caption{Left: Contour plots of interpolated stresses in y-direction in $MPa$. 
Right: Absolute errors of stresses in y-direction of solvers compared to reference solution at integration points in $MPa$. 
Top: Classical data-driven solver. 
Bottom: Extended data-driven solver by tensor voting.
Both computations performed on the same data \mbox{set 2} of size $16^3$.}
\label{fig:stresses2D}
\end{figure}

\noindent
As a second motivational example of the tensor voting extension to the data-driven paradigm, we consider a two-dimensional plane strain problem for a nonlinear and anisotropic elastic material.
In this case, the local phase space consists of pairs $(\feps,\fsigma)$ of strain and stress.
Since both quantities are symmetric second order tensors it follows that the corresponding phase space is six-dimensional.
For this simple example the dimensionality is high enough to raise questions about data set sampling and phase space coverage.
The smallest data set size of 100 points is certainly well feasible in the context of the truss example. For the plane strain case, however, it has to be taken into account that $100$ points in the 1D case would lead to $100^3$ data points in 2D. This is due to the fact that all three strain components (two normal strains and the shear strain) have to be covered. \\[0.3cm] 
We consider the problem of a plate with a hole under tensile loading $q(x)=200\,MPa$. The specimen is shown in Fig. \ref{fig:plateWithHole}.
Due to symmetry, the system can be reduced to one quarter (see Fig.\ \ref{fig:boundaryPWH}).
We assume non-linearly elastic material behavior. 
The stress is specified as
\begin{equation}
 \fsigma=\lambda g\big(\tr{\feps}\big)\fI + \mu \feps+\mathbb{D}\feps,
\end{equation}
where the function $g(x)=((|x|+a)^p-a^p)\sign{x}$ has been introduced. The parameters $a$ and $p$ are given by $a=0.001$ and $p=0.005$, respectively. Further, we need to define the Lam\'e constants $\lambda=57692,31\,MPa$ and $\mu=38461,54\,MPa$. 
The orthotropic elasticity matrix $\mathbb{D}$ is stated as
\begin{equation}
 \mathbb{D}=\sv C_{1111} & 2\nu(\bar\lambda+G_\perp) & 0\\ 2\nu(\bar\lambda+G_\perp) & \bar\lambda+2G_\perp & 0 \\ 0 & 0 & G_{||}\ev,
\end{equation}
where $C_{1111}=4.6875E$, $G_\perp=0.3E$, $G_{||}=0.2E$ and $\bar\lambda=(2\nu^2+1)/(15-20\nu^2)E$ are additional material parameters with $E=100000\,MPa$ and $\nu=0.3$.\\[0.3cm]
Next, we return to the question of convergence with respect to data set size.
We consider two different types of data distributions.
The first set contains data points randomly created within a range of $[-0.01, \, 0.01]$ for strains in each dimension (\mbox{set 1}).
The second set is created by a zero-mean normal distribution with a standard deviation of $0.005$ in all strain dimensions (\mbox{set 2}).
The error plots of 100 independent computations for \mbox{set 1} and \mbox{set 2} with varying set sizes are shown in \mbox{Fig. \ref{fig:convhistoideal2D}}.
Considering the results of the classical min-dist solution scheme, the results of \mbox{set 2} outperform those of \mbox{set 1} clearly.
This observation can already be classified as some kind of importance sampling.
Regarding \mbox{set 2}, more points are created close to the zero value, which seems to be advantageous for the problem under consideration.
The tensor voting solutions strictly outperform the results of the classical solution scheme, even if the relative variance of the solution error is higher, especially for \mbox{set 1}.\\[0.3cm]
The quality of the results of the tensor voting scheme based on the sparsest sampling size can well compete with the one of the classical solution scheme based on the finest data set sampling (see Fig.\  \ref{fig:convhistoideal2D}).
This is a highly satisfactory outcome for the suggested use of the tensor voting algorithm.
The absolute errors of the stresses in y-direction of the classical (min-dist) and enhanced (ten-vote) solver operating on $16^3$ data points are depicted in \mbox{Fig. \ref{fig:stresses2D}}.
It can be notified that the maximum stress error of the classical solver is about $250\,MPa$, whereas the maximum error of the tensor voting extension is only $15\,MPa$.
\subsection{Identification of phase space coverage of importance}
\begin{figure}[htbp]
\centering
\begin{subfigure}{0.49\textwidth}
\centering
\includegraphics[width=\textwidth]{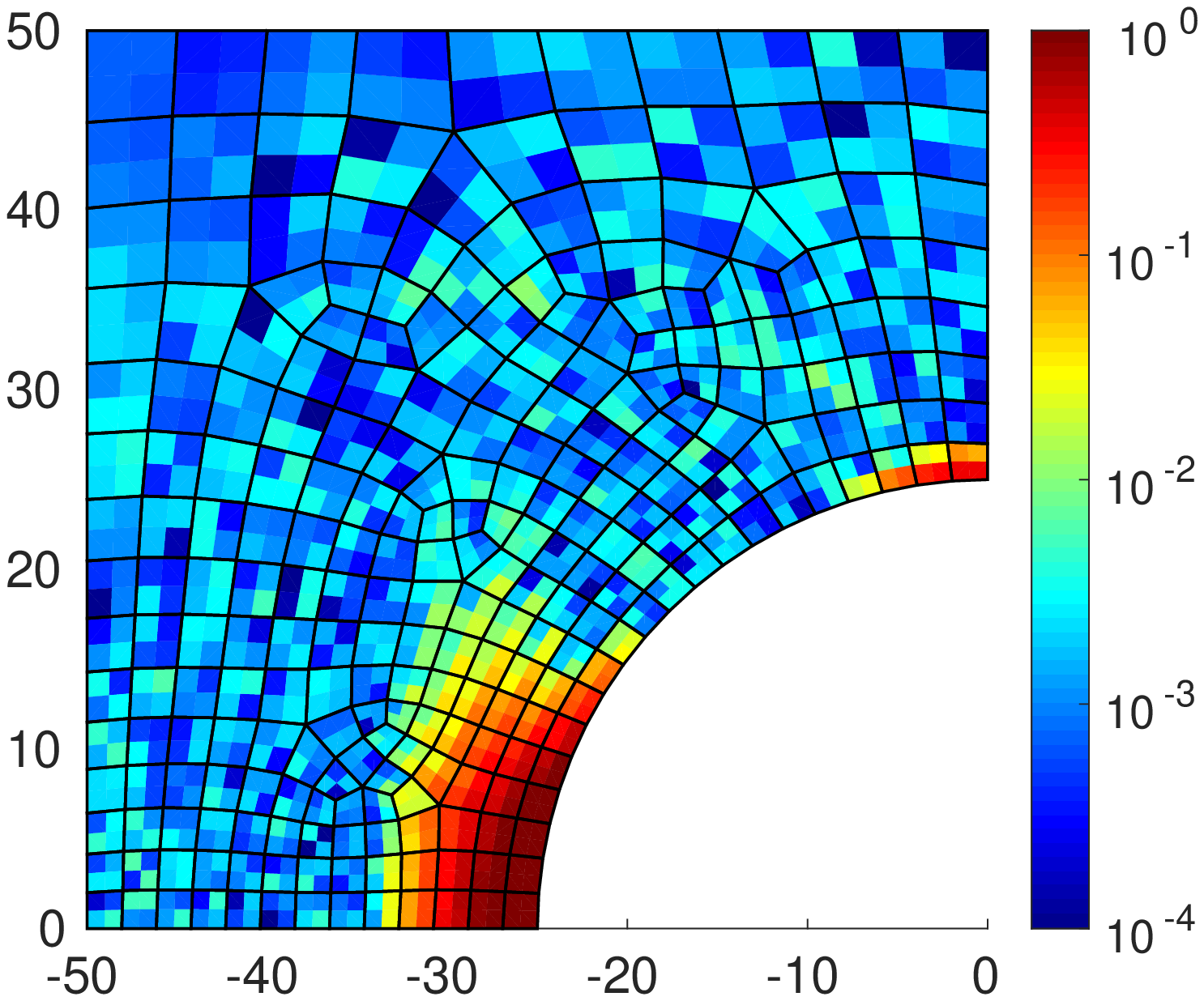}
\subcaption{}
\label{fig:phasespacecoverage1}
\end{subfigure}
\begin{subfigure}{0.49\textwidth}
\centering
\includegraphics[width=\textwidth]{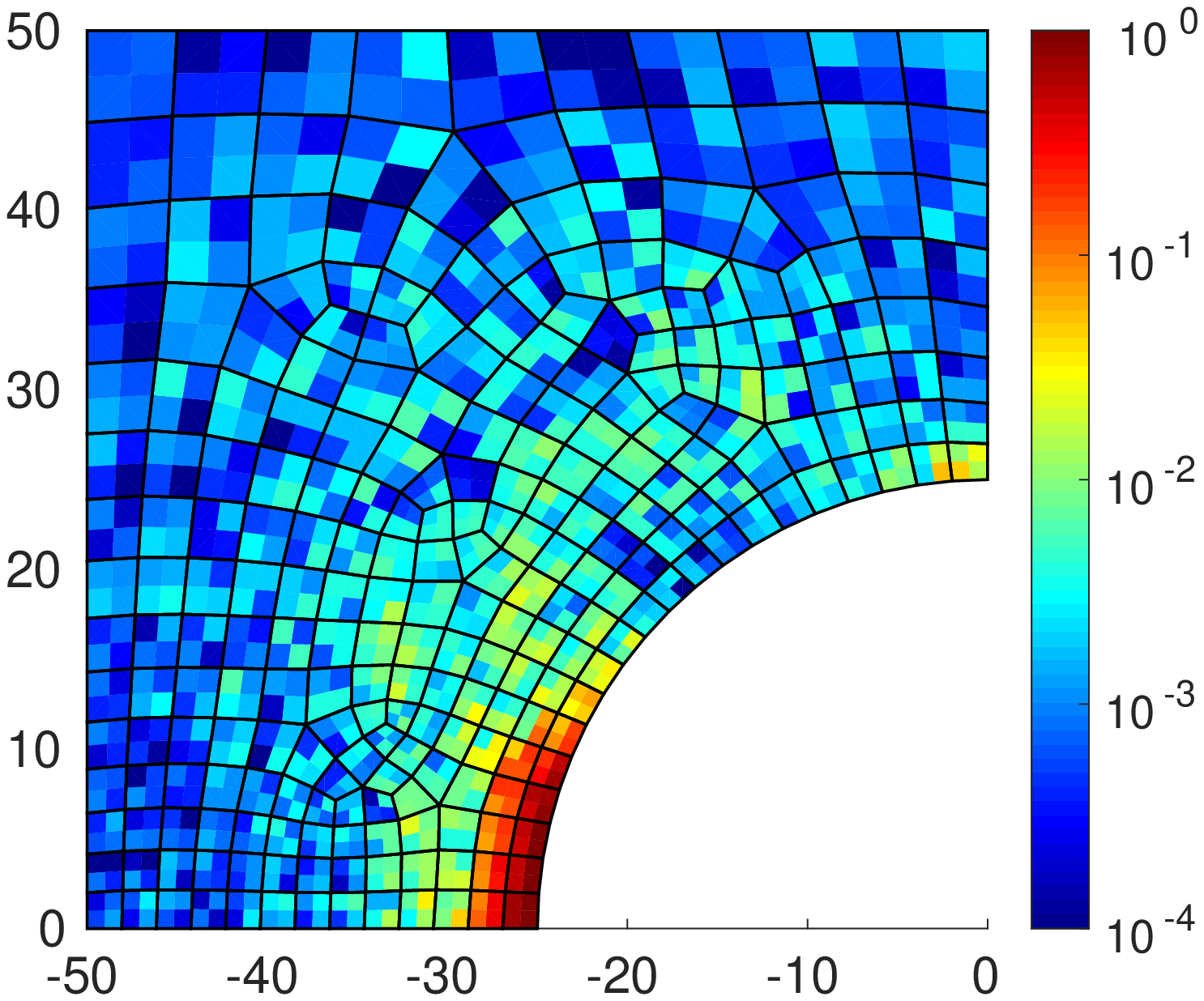}
\subcaption{}
\label{fig:phasespacecoverage2}
\end{subfigure}
\caption{Contour plots of local distance functions $d_e(\fz\e,\fy\e)$ at each Gaussian point after convergence of global iterations. 
Results for data sets of $32^3$ points. Left: data distribution of  \mbox{set 1}. Right: data distribution of \mbox{set 2}.}
\label{fig:phasespacecoverage}
\end{figure}

\noindent
Finally, we investigate the question of sample quality, i.e., the ability of a given data set to sample closely all the local states covered by the solution.
\mbox{Fig. \ref{fig:phasespacecoverage}} shows the contour plots of remaining local distances $d_e(\fz\e,\fy\e)$ after applying the min-dist tensor voting solver.
The solutions are computed with data sets created of $32^3$ points for \mbox{set 1} and \mbox{set 2}.
In contrast to the previous section, the strains of \mbox{set 1} are randomly sampled in a range of $[-0.025,\,0.025]$, whereas the standard deviation of \mbox{set 2} is chosen to be $0.025$.\\[0.3cm]
\mbox{Fig. \ref{fig:phasespacecoverage1}} shows that the phase space is not sampled adequately for the solution of the investigated boundary value problem.
This concluding remark follows by observing high values of the remaining distances in the regions where the highest strains and stresses are expected.
Thus, the corresponding and necessary data points with higher strains and stresses are not included in the data set which might lead to inaccurate results.
Considering the results obtained by the data distribution of \mbox{set 2} (see \mbox{Fig. \ref{fig:phasespacecoverage2}}), local distances are increasing continuously with increasing stresses and strains.
This behavior was expected since the density of the data set decreases systematically with increasing strains and stresses.\\[0.3cm]
In conclusion, we can state that by using the tensor voting extension it remains possible to identify regions of the phase space which are not covered adequately.
The analysis of the remaining local distances suggests a criterion for improving data sets adaptively.
A promising strategy is to identify the regions of phase space with the largest distances.
These regions can then be sampled more intensively by further testing.
Thus, the data set can be adaptively expanded in regions of high relevance to a particular application. 
Counter wise, the data set can be analyzed for irrelevant data to the problem.
These data points can be removed from the data set to improve the efficiency of the solver.
\section{Summary and concluding remarks} \label{sec:conclusion}
\noindent
We have formulated an extension to the model-free data-driven computing paradigm \citep{kirchdoerfer2016data}, which incorporates the use of pointwise tangent spaces. 
Hence, it is implied that the data has a specific underlying structure, which is used to receive additional information in the case of sparse data sets. 
The framework presented is a simple but effective plug-in method for the existing data-driven scheme. 
The proposed extension works in the framework of minimizing-distance as well as maximum-entropy data-driven computing.\\[0.3cm]
Tangent spaces are obtained through tensor voting \citep{mordohai2010dimensionality}, which is an eager machine learning technique and thus makes it possible to analyze the data in an offline step. 
The tensor voting method was recapitulated and simplified for the scope of application in the sense that we restricted the formulation to unoriented or rather ball tensors. 
Each data point assigns a tensor encoding the orientation of the underlying data structure.
Hereby, the orientation of a tensor is computed by accumulating votes from the point's nearest neighbors. 
Tangents and normals to the manifold are finally computed by means of spectral decomposition.\\[0.3cm]
The data-driven framework was extended by additionally minimizing the distance to a local tangent space. 
The procedure ensures to stay close to the data set but enables interpolation in regions of sparse data sampling. \\[0.3cm]
We have investigated the performance of the data-driven solver using the proposed extension with the aid of two particular examples of application, namely, the static equilibrium of nonlinear three-dimensional trusses and a 2D finite element discretized elastic solid. 
We showed by numerical testing that the initially proposed scheme is outperformed in the one-dimensional case even for noisy data. 
For noise-free data a convergence rate with respect to the data set size could be observed, which is twice as high as the classical data-driven solution scheme. 
For the elastic solid, it was shown that the performance of the new extension is of high performance even for very sparse data samplings.\\[0.3cm]
\textit{Data-driven computing paradigm.} The traditional phenomenological computing paradigm often faces problems fitting the data on which their solution is based on. 
Using data sets in computations directly removes this barrier and creates new possibilities in the arsenal of scientific computing.
As seen in this work, different mathematical objects like data sets with extended tangent spaces can be incorporated without facing convergence problems of the solver as in the classical computing paradigm.\\[0.3cm]
\textit{Higher-order data-driven schemes.} Due to the quadratic convergence rate with respect to the data set size we can classify the proposed method as a second-order data-driven scheme.
This higher-order convergence rate goes with the requirement of a data set of a certain regularity.
In the case of discontinuities or non-smooth behavior in the data set the method might face problems.
To reduce possible failures, a maximal distance between the original data and the points on the tangent spaces could be defined.  \\[0.3cm]
\textit{Connection to Machine learning.} The final goal of the data-driven computing paradigm is to use a data set at its entity without neglecting any relevant information.
In this work we go one step further in the sense that we want to understand how data points are related and oriented to each other. 
Hence, tangent spaces are assigned to each data point representing the locally linear structure of the data set.
It should be pointed out that the tensor voting method provides even more information than used in this work. 
The use of stick and generic votes, as well as the analysis of eigenvalues during the voting process, can provide further improvement and automation.\\[0.3cm]
\textit{Data acquisition.} In the work presented here, we have assumed that the data is merely available in a specific density and quality. 
In reality,  we face the critical issue of material data set acquisition, especially for higher-dimensional cases. 
The main idea to overcome this problem is the use of importance sampling techniques. 
These methods generate data sets that are highly relevant to the particular problem under consideration. 
The self-consistent data-driven identification approach of Leygue et al. \citep{leygue2018data} was introduced to create such goal-oriented data sets. 
In \citep{stainier2019model} it has already been shown, that utilizing importance sampling the performance can be significantly increased.
The combination of data acquisition and data-driven computing suggest important directions for further research.
\section*{Acknowledgments}
\noindent
MO gratefully acknowledges the support of the Deutsche Forschungsgemeinschaft (DFG) through the Sonderforschungsbereich 1060 ``The mathematics of emergent effects". 
SR and RE gratefully acknowledge the  financial support of the Deutsche Forschungsgemeinschaft (DFG) through the project RE 1057/40-2 ``Model order reduction in space and parameter dimension — towards damage-based modeling of polymorphic uncertainty in the context of robustness and reliability" within the priority program SPP 1886 ``Polymorphic uncertainty modelling for the numerical design of structures". 
Further, SR and RE rewardingly acknowledge the funding by the Excellence Initiative of the German federal and state governments through the project ``Predictive Hierarchical Simulation".
Finally, all authors acknowledge the financial support of the DFG and French Agence Nationale de la Recherche (ANR) through the project ``Direct Data-Driven Computational Mechanics for Anelastic Material Behaviours" (project numbers: ANR-19-CE46-0012-01, RE 1057/47-1) within the French-German Collaboration for Joint Projects in Natural, Life and Engineering (NLE) Sciences.

\bibliography{TensorVotingBib}

\end{document}